\documentclass[a4paper,11pt]{article}
\pdfoutput=1 % if your are submitting a pdflatex (i.e. if you have
\usepackage{jcappub} % for details on the use of the package, please
\usepackage[T1]{fontenc} % if needed

\usepackage{orcidlink, pgfplots, booktabs, bm, subcaption, multirow}

\usepgfplotslibrary{groupplots}
\pgfplotsset{compat=1.18}
 
\definecolor{responsecolor}{rgb}{0,0,0}%{1,0,0}

\pgfmathdeclarefunction{gamma}{1}{\pgfmathparse{exp(ln(2.5066282746310002) + (#1-0.5)*ln(#1) - #1 + 1/(12*#1) - 1/(360*#1^3))}}

\title{\boldmath The Ellis and Baldwin test of the Cosmic Dipole:
Exploring the impact of multiple flux density cuts}

\author[1]{Vasudev Mittal\note{Corresponding author} \orcidlink{0000-0002-1708-6088}}
\author{and Geraint F. Lewis \orcidlink{0000-0003-3081-9319}}

\affiliation{Sydney Institute for Astronomy, School of Physics A28\\ The University of Sydney, NSW 2006, Australia}

\emailAdd{vasudev.mittal@sydney.edu.au}

\abstract{
        The cosmic dipole tension - the discrepancy between the Cosmic Microwave Background kinematic dipole and the matter dipole inferred from all-sky surveys poses a significant challenge to the Cosmological Principle, which dictates that the universe is homogeneous and isotropic at the largest scales.
        Traditional measurement of the matter dipole requires selecting an appropriate limiting flux and calculating the dipolar modulation using sources brighter than the flux.
        This approach, however, ignores the shape of the source luminosity function (LF) and deprives the analysis of this crucial information.
        In this study, we present a new approach to calculate the matter dipole by integrating the source flux distribution into the analysis.
        We achieve this by dividing the catalogue into disjoint flux bins and simultaneously fitting the matter dipole across them.
        For non-power-law LFs, this method gives a higher Bayes factor - and hence a better description of the matter dipole - as compared to the traditional approach.
        The method works best when the flux cuts are selected in regions where the LF's shape changes significantly.
        We discuss the feasibility of this method for upcoming cosmological surveys and show that it has the potential to yield decisive results at both radio and infrared wavelengths.
}

\keywords{Bayesian reasoning, cosmological parameters from LSS, galaxy surveys}

\begin{document}
\maketitle

\flushbottom

\section{Introduction}
The foundations of modern cosmology are built on the assumption that the universe is spatially maximally symmetric, i.e. homogeneous and isotropic at the largest scales.
Although Einstein first used this assumption to describe a static universe within General Relativity, it was Milne who established it as the `Cosmological Principle' (CP) \citep{milne1935}.
Today, CP is used to derive the Friedmann–Lemaître–Robertson–Walker (FLRW) metric, which, when used in conjunction with Einstein's equations, underpins the current standard model of cosmology -- the $\Lambda \text{CDM}$ model \citep{Weinberg:2008zzc}.
However, the notion that our universe obeys the CP is fundamentally an \emph{a priori} assumption, originally introduced without direct observational support. 
This makes it essential to establish it \emph{a posteriori}, grounded in robust astronomical evidence.
Over the past century, a number of independent astronomical probes have supported the CP, but the most substantial evidence in its favour comes from the smoothness of the Cosmic Microwave Background's (CMB) temperature map.
Small-scale thermal fluctuations in this map, which are of the order of $\mathcal{O}(10\mu K)$, are theorised to be the seeds of the cosmic structure.
These anisotropies, however, are subdominant to the $mK$ scale dipole anisotropy, which is interpreted to arise due to our motion with respect to the `Cosmic Rest Frame'.
The kinematic interpretation of this dipole, inferred solely from CMB observations \citep{aluri2023}, therefore requires independent verification using independent observables.
If the ansatz of the observer's motion is correct, then the all-sky distribution of high-redshift sources should have a similar dipole imprinted upon it \citep{ellis1984}.
This `matter dipole', which has been observed in all-sky distributions of both radio-galaxies \citep{blake2002a} and quasars \citep{secrest2021}, disagrees with its CMB counterpart.
The disagreement, often dubbed the `dipole tension', serves as a test of the assumption of isotropy (popularly known as the Ellis \& Baldwin test) and is a significant problem in modern cosmology \citep{Peebles_2022, Secrest2025}.

A key assumption in Ellis \& Baldwin's test is that the observer can see the universe up to a limiting flux-density.
Under this, the test measures the change in the source density when the observer shifts from a stationary frame to one moving with a velocity $v$.
Hence, the amplitude of the matter dipole depends on the flux limit of the observations, and selecting an appropriate flux limit is a key step in cosmic dipole studies.
In practice, this selection is largely a matter of statistics: pushing to fainter flux limits increases the number of detected sources and thereby improves the signal-to-noise of the dipole estimate.
It is not uncommon for studies to establish the validity of their results by demonstrating consistency in the matter dipole estimated across different choices of limiting flux \citep[see, for example,][]{siewert2021}.
This approach assumes that each flux-limited version of the dataset is statistically independent, and that the underlying dipole-generating effect is flux-dependent.\footnote{In this study, we use the phrases `catalogue with limiting flux' and `flux-limited catalogue' interchangeably.
We also refer to `flux-density' simply as `flux'.}
However, for the same dataset, a catalogue with a brighter limiting flux is a subset of its fainter counterpart, and under the E\&B test, the matter dipole is generated by the observer's motion.
This challenges the above assumptions and highlights avenues for improving the methodology.

In this study, we address this point and discuss a method for studying the catalogue across multiple flux limits. 
We divide a dataset into multiple disjoint flux bins, and then use Bayesian statistics to jointly analyse all flux bins to calculate the matter dipole.
We select our parameter space by noting that the dipole-generating effect is independent of the flux.
So, the matter dipole at different limiting fluxes should point in the same direction, and the amplitudes should be proportional to each other.
We demonstrate the efficacy of our method by using it to study the matter dipole in simulated catalogues.

This paper is organised as follows.
The next Section discusses the relevant literature background.
Then we discuss the setup used to generate the simulated catalogues in Section ~\ref{sec: simulated skies}.
In Section ~\ref{sec: Bayesian framework}, we discuss the Bayesian framework used to calculate the matter dipole.
We present our results in Section ~\ref{sec: results} and discuss their properties and implications in Section ~\ref{sec: discussion}.

\section{Background}\label{sec: background}
Special relativity dictates that, under the kinematic interpretation, the dipolar modulations will be generated by two effects: relativistic aberration, which shrinks the observed solid angles in the direction of motion, and the Doppler shift of light, which modifies the observed source flux/observed CMB temperature based on their relative location with respect to the direction of motion. 
Taking these into account, the CMB's $mK$ scale dipole predicts that the solar system is moving towards the Galactic coordinates $(264{.}^{\circ}021, 48{.}^{\circ}253)$ with a velocity of $369.82 \pm 0.11 \,\text{km}\,\text{s}^{-1}$ \citep{planck2020}.
Ellis \& Baldwin's work \citep{ellis1984} calculates the matter dipole observed in the all-sky distribution of sources by making the following additional assumption about their population.
\begin{itemize}
    \item the sources should be at high redshift: the mean $z$ of the population should be $\approx 1$ \citep{ellis1984};
    \item the Spectral Energy Distribution (SED) has a power law dependence on the observation frequency, i.e. $S \propto f^{-\alpha}$;
    \item the flux distribution obeys a power-law survival function in the vicinity of the limiting flux: $N(>S_{\texttt{lim}}) \propto S_{\texttt{lim}}^{-x}$.
\end{itemize}
Then, the number density contrast in the direction $\bf n$ due to the observer's velocity $\bf v$ is
\begin{equation}\label{eq:ellis}
    \frac{\Delta N ({\bf n})}{N} = [2+x(1+\alpha)] \frac{\bf v}{c}\cdot{\bf n}.
\end{equation}
For a velocity of $\sim 400 \,\text{km}\,\text{s}^{-1}$, this is a $\mathcal {O}(0.1\%)$ effect, and hence needs $\sim\mathcal {O}(10^5)$ sources for a significant detection \citep{ellis1984}.
Over the years, a number of studies have calculated the matter dipole in different catalogues of radio galaxies, IR and optical quasars, and the current consensus is that the matter dipole aligns with the kinematic dipole's direction, but the former has a significantly higher amplitude \citep[see for eg][]{2002Natur.416..150B, crawford_2009, singal2011, rubart2013,colin2017, bengaly2018,secrest2021, 2022ApJ...937L..31S, darling2022, dam2023, wagenveld2023, mittal2024, Oayda:2024hnu, Bohme:2025nvu, vonHausegger:2025iuo}.
An in-depth discussion of the studies is also available in some review papers \citep{aluri2023, Secrest:2025wyu}.

In addition to studying the catalogues, several works have also suggested improvements to the E\&B methodology. 
% We discuss some notable ones here.
Some studies \citep{dalang2022} have proposed that the spectral index, $\alpha$ and the population's shape parameter, $x$, should have a redshift dependence, but it has been shown that the inferred matter dipole is robust against a redshift-dependent evolution \citep{vonHausegger:2024jan}.
It has also been suggested that $\alpha$ and $x$ should be calculated for a small subset of the population around the flux limit \citep{vonHausegger:2024jan}.
Some outstanding issues related to leakage from higher multipoles into the dipole have been addressed by \citep{Oayda:2024voo}.
Other studies have focused on streamlining existing statistical frameworks \citep{mittal2024} and on implementing new statistical tools \citep{Oayda:2026afy, Land-Strykowski:2025gkz}.
Theoretical formulations for extending E\&B's test to a general population of astronomical sources have been discussed in some recent studies \citep{Bonnefous:2026hpe, Takeuchi:2026can}.
Finally, studies have discussed methods for testing dipole anisotropy using previously unexplored astronomical observables \citep{Blumke:2025nrq, Martin:2025ywz, Yasin:2026nte, Millon:2026nwo}.
Even though these improvements have helped in relaxing the assumptions on which E\&B's test is based, the philosophy behind the calculation of the matter dipole remains unchanged.

One fundamental point that has not been adequately addressed in the literature concerns the selection of an appropriate flux limit for calculating the matter dipole.
Since the amplitude of the matter dipole depends on the cumulative population and SED of sources near the flux limit, selecting a reasonable flux limit is the key to a proper inference.
A broad consensus on this issue is as follows: the catalogue should have high completeness and reliability beyond the limiting flux, the impact of instrumental biases should be minimised, and the source density should be homogenised across the sky.
Once such a threshold is identified, it is equally plausible to select any brighter flux value as the limiting flux.
Many studies have utilised this concept and demonstrated that their results are consistent across a number of limiting fluxes \citep{blake2002a, rubart2013, tiwari2016, siewert2021}.
These studies further show that the computed matter dipoles for various limiting fluxes are consistent with one another, demonstrating that their conclusions are not contaminated by any flux-dependent effect.

This method, however, glosses over a number of statistical and physical nuances.
First, a critical assumption in demonstrating consistency is that the different flux-limited datasets are statistically independent. 
Since a catalogue with a brighter limiting flux is a subset of one with a fainter limiting flux, the former is contained within the latter. 
Consequently, the dipole estimate in the latter already incorporates information from the former. 
This allows us to claim only internal consistency, which is a weaker statement than external or independent consistency.
Second, when we calculate the matter dipole for different flux-limited versions in isolation, we implicitly assume that the dipole at each flux limit arises from a different (flux-dependent) underlying cause. 
However, the fundamental premise of E\&B’s test is that the dipolar modulation is induced by the observer’s motion, which is a flux-independent effect.
These subtleties point toward a refinement of the methodology. 
A more plausible approach is to first construct disjoint, flux-binned versions of the dataset, thereby ensuring statistical independence, and then fit a dipole at each flux limit, assuming it points in the same direction across all flux limits while allowing its amplitude to depend on the source population near each flux limit.

In this work, we demonstrate the power of this improved statistical approach by calculating the matter dipole for simulated all-sky catalogues.
We construct multiple flux-limited versions of a catalogue, use them to define disjoint flux bins, and then perform a Bayesian analysis to jointly analyse all versions in order to estimate the amplitude and direction of the matter dipole.
Since the matter dipole's amplitude depends on the properties of the source population near the flux limit, we will study simulated populations having different luminosity functions.
This would help establish the efficacy of our method for analysing source populations with a complex distribution of sources around the limiting flux. 
We will also use different values of the spectral indices to demonstrate that the analysis is independent of the shape of the population's SED.

\section{Mock Catalogues}\label{sec: simulated skies}
\subsection{Simulated Skies}
To construct the mock catalogues, we first generate an isotropic distribution of $N$ sources scattered over the celestial sphere and assign them a flux value sampled from a Luminosity Function (LF).
We then shift the observer to a reference frame moving with a velocity $= 370\, \text{km}\,\text{s}^{-1}$ in the CMB dipole direction. 
This shift is achieved by modulating the coordinates and flux of each source to account for relativistic aberration and Doppler boosting, respectively.

\begin{figure*}[htbp]
\centering
\resizebox{\textwidth}{!}{
\begin{tikzpicture}
\begin{groupplot}[group style={
        group size=3 by 1,
        horizontal sep=0.5cm
    },
    width=0.3\textwidth,
    height=4.5cm,
    scale only axis,
    enlarge x limits=false,
    enlarge y limits=false,
    domain=9:15000,
    samples=400,
    xmode=log,
    ymode=log,
    tick align=inside,
    group/xticklabels at=edge bottom, % shared x ticks at bottom
    legend style={at={(0.03,0.07)},anchor=south west, inner sep=1pt}
]

\nextgroupplot[
    ylabel={},
    yticklabels={},
    title=Power-law
]

\addplot[blue!70!black, dash dot]   {0.7*(10^0.7)/(x^1.7)};
\addplot[purple!70!black, dash dot dot] {0.9*(10^0.9)/(x^1.9)};
\addplot[yellow!80!orange, dashed] {1.1*(10^1.1)/(x^2.1)};
\addplot[green!60!black, dotted] {1.3*(10^1.3)/(x^2.3)};
\legend{$x=0.7$,$x=0.9$,$x=1.1$,$x=1.3$}

\nextgroupplot[
    yticklabels={},
    title=Schechter
]

\pgfmathsetmacro{\xstar}{600}

\addplot[blue!70!black, dash dot]   {((x/\xstar)^(-0.7))*exp(-x/\xstar)/2.991};
\addplot[purple!70!black, dash dot dot] {((x/\xstar)^(-0.9))*exp(-x/\xstar)/9.513};
\addplot[yellow!80!orange, dashed] {((x/\xstar)^(-1.1))*exp(-x/\xstar)/10.686};
\addplot[green!60!black, dotted] {((x/\xstar)^(-1.3))*exp(-x/\xstar)/4.326};
\legend{$\upsilon=0.7$,$\upsilon=0.9$,$\upsilon=1.1$,$\upsilon=1.3$}

\nextgroupplot[
    yticklabels={},
    title=Double power-law
]

\def\Sm{10}    % x_min
\def\St{200}   % break
\def\xone{0.7} % slope 1

\def\xTwoA{1.4}
\def\xTwoB{1.6}
\def\xTwoC{1.8}

% C1 arbitrary, just for plotting
\def\COne{1.0}

% Continuity factors
\pgfmathsetmacro{\CTwoA}{\COne * (\St)^( \xTwoA - \xone )}
\pgfmathsetmacro{\CTwoB}{\COne * (\St)^( \xTwoB - \xone )}
\pgfmathsetmacro{\CTwoC}{\COne * (\St)^( \xTwoC - \xone )}

\addplot[blue, thick, dash dot, domain=10:200]{\COne * x^(-1-\xone)};
\addplot[purple, thick, dash dot dot, domain=10:200]{\COne * x^(-1-\xone)};
\addplot[green, thick, dotted, domain=10:200]{\COne * x^(-1-\xone)};

\addplot[blue, thick, dash dot, domain=200:10000]{\CTwoA * x^(-1-\xTwoA)};
\addplot[purple, thick, dash dot dot, domain=200:10000]{\CTwoB * x^(-1-\xTwoB)};
\addplot[green, thick, dotted, domain=200:10000]{\CTwoC * x^(-1-\xTwoC)};

\legend{,,,$x_2=1.4$,$x_2=1.6$,$x_2=1.8$}

\end{groupplot}
\node at ($(group c1r1.south west)!0.5!(group c3r1.south east) + (0,-0.9cm)$) {{Flux (mJy)}};
\node[rotate=90, anchor=center] at ($(group c1r1.north west)!0.5!(group c1r1.south west) + (-0.6cm,0)$) {{PDF}};

\end{tikzpicture}
}
\caption{Some of the luminosity functions used in this work, visualised on logarithmic plots.}
\label{fig: pdfs}
\end{figure*}
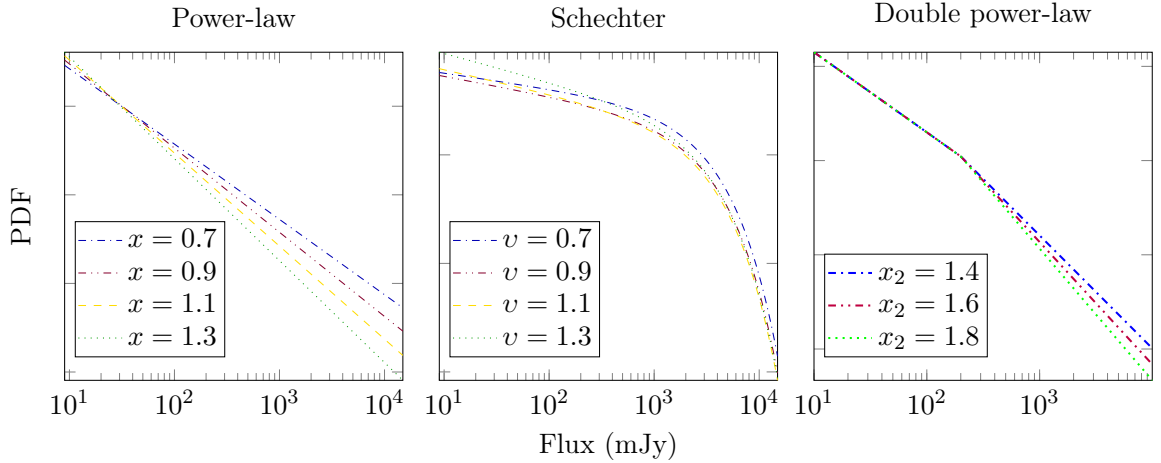

We start by sampling 50 million isotropically scattered sources to generate our initial distribution.
This high initial source count gives us a sizable number of sources in the final catalogues and enables us to probe the impact of brighter flux cuts.\footnote{We discuss the impact of source counts in the discussion section.}
The next step is to select an appropriate LF to sample source fluxes.
E\&B's test does not depend on any specific choice of the flux distribution; it only assumes that the cumulative source count (per unit solid angle) above the flux limit is approximated by the power-law survival function evaluated at the limiting flux.
The simplest flux distribution satisfying this requirement is the power-law function, which has been used in numerous important studies to assign the fluxes to a  simulated source populations \citep[see, for example,][]{2022ApJ...937L..31S, wagenveld2023}.
This distribution gives a constant value of $x$ %(or equivalently, the dipole amplitude considering a constant $\alpha$) 
for any limiting flux.
However, a priori, there is no reason why the flux distribution of a source population should follow a power-law.
Therefore, in this work, in addition to the power-law function, we work with two other popular LFs: the double power-law and the Schechter/gamma function.
Their functional forms, free parameters, and the magnitudes of the free parameters we use in this study are listed in Table \ref{tab: pdfs}, while their shapes for some choices of free parameters are shown in Figure \ref{fig: pdfs}.
When sampling from the luminosity functions, we adopt a minimum flux of $10\ mJy$; this choice is arbitrary and does not affect our analysis.

\begin{table}[htbp]
\centering
\resizebox{\textwidth}{!}{
\begin{tabular}{c c c c}
\toprule
Flux Distribution & PDF, $N(S)$ & Free Parameters & Free Parameter values \\
\midrule
\multirow{2}{*}{Power-law} 
& \multirow{2}{*}{$\displaystyle \frac{xS_{m}^{x}}{S^{x+1}},\quad S>S_m$}
& $x$ = slope of the survival function
& $x = \{0.7,0.8,0.9,1,1.1,1.2,1.3,1.4,1.5\}$ \\
& & $S_{m}$ = minimum flux value & $S_{m} = 10\,\mathrm{mJy}$ \\
\midrule
\multirow{4}{*}{Double power-law}
& \multirow{4}{*}{\centering
\large$\displaystyle
\begin{array}{l}
C_1 S^{-(x_{1}+1)}, \quad S\in [S_m,S_t] \\
C_2 S^{-(x_{2}+1)}, \quad S > S_t
\end{array}$}
& $x_i$ = slopes of the survival function 
& $x_1 = 0.7,\quad x_2 = \{1.4,1.5,1.6,1.7,1.8\}$ \\
& & $S_{m}$ = minimum flux value & $S_{m} = 10\,\mathrm{mJy}$ \\
& & $S_{t}$ = transition flux & $S_{t} = 200\,\mathrm{mJy}$ \\
& & $C_i$ = normalizations & $C_i$ determined numerically \\
\midrule
\multirow{3}{*}{Schechter}
& \multirow{3}{*}{$\displaystyle \frac{e^{-S/S_{*}}}{S_{*}\Gamma(1-\upsilon)}
\left(\frac{S}{S_{*}}\right)^{-\upsilon}, \quad S>S_m$}
& $\upsilon$ = slope of the PDF
& $\upsilon = \{0.7,0.8,0.9,1.1,1.2,1.3,1.4,1.5\}$ \\
& & $S_{*}$ = exponential cutoff scale & $S_{*} = 400\,\mathrm{mJy}$ \\
& & $S_{m}$ = minimum flux value & $S_{m} = 10\,\mathrm{mJy}$ \\
\bottomrule
\end{tabular}}
\caption{List of the luminosity functions used in this study. Their functional forms, free parameters, and parameter values are listed. Each set of parameter values defines a distinct realisation of the corresponding flux distribution.}
\label{tab: pdfs}
\end{table}

Next, we need to assign the spectral index $\alpha$ to the source population.
This parameter impacts the contribution of Doppler boosting to the matter dipole's amplitude.
For ease of analysis, we assign a uniform spectral index to the entire source population.
We select the spectral index from $\alpha \in \{0.7, 0.9, 1.1\}$ to investigate its magnitude's impact on the analysis. 
{\color{responsecolor} Our choice of $\alpha$ is representative of the range of spectral indices of extragalactic sources, with radio surveys like NVSS having a mean $\alpha = 0.75$ \citep{Oayda:2024hnu}, while infrared surveys like CatWISE have a mean $\alpha = 1.26$ \citep{secrest2021}.}

After selecting these parameters, we construct $30$ isotropic skies for each $(LF,\alpha)$ combination.
We then modulate the position and flux of each source to account for the observer's motion.
After generating the modulated catalogues, we select all sources with flux $>12\ mJy$, as this removes edge effects from the modulated luminosity function.
This gives us the mock catalogues, which we will use for analysis.

\subsection{Pixelising the sky}
E\&B's test tracks changes in source density in any direction caused by the observer's motion.
This necessitates discretising the sky into a finite set of equal-area patches.
Ideally, each patch's spatial extent should be small enough to be considered as a point.
At the same time, it should have sufficient sources to make the shot-noise subdominant to the mean density of the catalogue. 
We fulfil these requirements by pixelating the sky using the \textsc{healpix}\footnote{\url{https://healpix.sourceforge.io/}} algorithm in \textsc{Python}, through the \textsc{healpy} package \citep{Gorski2005, Zonca2019}.
We use \texttt{nside=64} to generate $49,152$ equal-area pixels, each having an area of $\approx 0.83\ deg^2$.
We then assign a number density and a list of fluxes to each pixel, accounting for all sources within its spatial extent. 
We use this list of fluxes to make the flux-binned sky maps, where the source count in each pixel is the number of entries in the list within the limits of the bin.

\subsection{Expected matter dipole amplitude}
Before proceeding further, we need a method to calculate the amplitude of the matter dipole, which we have injected in the catalogue using aberration and Doppler boosting. 
This will be used to check whether the Bayesian inference yields the correct dipole as an output. 
We calculate the expected dipole amplitude at the flux limit without assuming any specific shape of the cumulative number counts.
This is done by tracking changes in source counts due to the observer's motion.
Although the method has been introduced by us in one of our previous works \citep{mittal2024}, we discuss it briefly here as well:
\begin{enumerate}
    \item We first compute the number of sources beyond a flux threshold $S_l$, and denote them with $n_i$
    \item We apply Doppler boost to the j-th source, by modulating its flux from $S_j$ to $ S_j \delta^{1 + \alpha}$, where $\delta = \gamma (1 + \frac{v}{c} \cos \theta)$ and $\gamma$ is the Lorentz factor.
    \item Then, we count the number of sources with boosted flux beyond the threshold, and multiply this number by $\delta^2$ to account for relativistic aberration. 
    We denote this by $n_b$
    \item Using this, we calculate the expected dipole amplitude at the flux threshold as 
    \begin{equation}
        \mathcal{D} = \frac{n_b - n_i}{n_i}.
    \end{equation} 
\end{enumerate}
Note that we measure the amplitude along the line of motion, where the density enhancement is maximum, and set $\theta=0$, which gives $\delta = \gamma (1 + \frac{v}{c})$.

\section{Bayesian Framework}\label{sec: Bayesian framework}
\subsection{Likelihood function}
In this study, we will use Bayesian analysis to analyse the matter dipole present in the mock catalogues.
We use Bayes's theorem 
\begin{equation}
    P( \Theta \, | \, \mathbf{D}, M )
        = \frac{\mathcal{L}(\mathbf{D} \, | \, \Theta, M) \pi(\Theta \, | \, M)}
        {\mathcal{Z}(\mathbf{D} \, | \, M)}
        \label{eq:bayes-theorem}
\end{equation}
to determine the probability distribution of a model $M$'s free parameters $\Theta$, if we fit it to a dataset $\mathbf{D}$.
This allows us to incorporate our prior belief about $\Theta$'s in the analysis pipeline using the prior probability function $\pi(\Theta \, | \, M)$.
The denominator $\mathcal{Z}$ is dubbed as the marginal likelihood and is a normalisation over the parameter space $\mathcal{Z} = \int_{\Omega_\Theta} \mathcal{L}(\mathbf{D} \, | \, \Theta, M) \pi(\Theta \, | \, M) d\Theta$.
It is used to compare competing hypotheses by constructing the Bayes factors
\begin{equation}
    \ln B_{ij} = \ln \mathcal{Z}_i - \ln \mathcal{Z}_j
\end{equation}
We use Jeffreys's scale \citep{kass1995} to qualitatively interpret the Bayes factors.

To specify a likelihood function, we note that in a flux-limited dataset, the counting statistics in each pixel follow a Poisson distribution - the observed source count in a pixel has a shot noise contribution arising from sky pixelisation.
Thus, we model the likelihood function by assuming that the mean source count in the i-th pixel is characterised by a rate parameter $\lambda_i$, and the probability of observing $\mathcal{N}_i$ sources is defined by a Poisson distribution taking $\lambda_i$ as the input:
\begin{equation}
    % \mathcal{L}(\mathcal{N}_i \, | \, \Theta, M) = \frac{e^{-\lambda_i} \lambda_i^{\mathcal{N}_i}}{\mathcal{N}_i !} \Rightarrow
     \mathcal{L}(\mathbf{D} \, | \, \Theta, M) = \prod_{i = 1}^{n_{\texttt{pix}}} \frac{e^{-\lambda_i} \lambda_i^{\mathcal{N}_i}}{\mathcal{N}_i !}
     \label{eq: likelihood single}
\end{equation}
This likelihood function has been used by previous studies \citep{dam2023, mittal2024} to estimate dipole parameters from all-sky quasar datasets.
We need to adapt this likelihood function for the joint analysis of the mock catalogues.

To achieve this, the first idea is to create a joint likelihood by multiplying the likelihoods for different flux-limited versions of the mock catalogue.
But this idea runs into the effects related to overcounting.
Let's say we have two flux limited catalogues with limiting fluxes $S_{1}$ and $S_{2}$, such that $S_{1}>S_{2}$, then $\mathcal{N}(>S_1)$ will have a contribution from $\mathcal{N}(>S_2)$.
Consequently, the brighter sources will contribute to both $\mathcal{L}_{S_{1}}$ and $\mathcal{L}_{S_{2}}$, which {\color{responsecolor} violates} the statistical independence of individual likelihoods.
% will increase their imprint on the inferred matter dipole.
To counter this, we need to ensure that each source in the dataset only contributes once to the net likelihood function.
A natural way to achieve this is to divide the catalogue into multiple disjoint flux bins and jointly analyse them to determine the matter dipole.
This gives us the following likelihood function:
\begin{equation}\label{eq: likelihood_joint}
     \mathcal{L}_{\texttt{net}}(\mathbf{D} \, | \, \Theta, M) = \prod_{j = 1}^{n_\texttt{bins}}\left(\prod_{i = 1}^{n_{\texttt{pix}}} \frac{e^{-\lambda_{i,j}} \lambda_{i,j}^{\mathcal{N}_{i,j}}}{\mathcal{N}_{i,j} !}\right)
\end{equation}
where $\mathcal{N}_{i,j}$ denotes the number of sources present in, and $\lambda_{i,j}$ is the rate parameter for the i-th pixel and j-th flux bin.
We calculate $\mathcal{N}_{i,j}$ by taking the difference of the $\mathcal{N}(>S)$ at the upper and lower flux limits of the catalogue, i.e. $\mathcal{N}_{i,j} = \mathcal{N}(>S_{\texttt{lower}}) - \mathcal{N}(>S_{\texttt{upper}})$.
Similarly, we specify the rate parameter as $\lambda_{i,j} = \lambda_{\texttt{lower}}-\lambda_{\texttt{upper}}$, where $\lambda_{\texttt{lower}}$ and $\lambda_{\texttt{upper}}$ are the rate parameters at the flux limits.
This likelihood removes the excess contribution of the brighter sources to the likelihood function.

We calculate the posterior distributions and Bayes factors using the Nested Sampling algorithm, which samples posteriors in shells of increasing likelihood \citep{skilling2004, skilling2006}.
We use a Nested Sampling Monte Carlo algorithm \textsc{MLFriends} \citep{2016S&C....26..383B, 2019PASP..131j8005B}, which efficiently samples likelihood shells for complex multi-dimensional likelihood space.
This algorithm has been implemented in the \textsc{UltraNest}\footnote{\url{https://johannesbuchner.github.io/UltraNest/}} \textsc{python} package \citep{2021JOSS....6.3001B}.

\subsection{Hypotheses and Priors}
\subsubsection{Dipole}
Our aim is to determine the matter dipole present in the catalogue by jointly analysing the disjoint flux bins.
This requires us to write a functional form of the expected source count, i.e. the rate parameter $\lambda_{i,j}$ for the flux bins.
As we have discussed above, the rate parameter for a flux bin can be written in terms of the rate parameters at the flux limits. 
Similar to our previous works \citep{dam2023, mittal2024}, we define the expected source counts by noting that for an observer in motion, the observed source count in any pixel of a flux-limited catalogue is a dipole modulated isotropic source count.
In other words, for a catalogue having sources up to the faintness $S_{\texttt{lim}}$, the rate parameter for the i-th pixel is
\begin{equation}
    \lambda_{i} = \bar{N}(>S_{\texttt{lim}})(1+\mathcal{D}_{S_{\texttt{lim}}} cos\theta_i)
\end{equation}
where $\theta_i$ is the angular offset between the expected dipole direction $(l^\circ, b^\circ)$ and the pixel coordinates; $\mathcal{D}_{S_{\texttt{lim}}}$ is the dipole amplitude at the flux limit, and $\bar{N}(>S_{\texttt{lim}})$ is the mean count of the isotropic sky.
Using this, we define $\lambda_{i,j}$ as
\begin{equation}
    \lambda_{i,j} = \bar{N}(>S_{j,l})(1+\mathcal{D}_{S_{j,l}} cos\theta_i) - \bar{N}(>S_{j,u})(1+\mathcal{D}_{S_{j,u}} cos\theta_i)
\end{equation}
where $S_{j,l}$ and $S_{j,u}$ are the lower and upper limits of the flux bins.
Since the flux bins are disjoint, we have $S_{j+1,l}=S_{j,u}$ and our parameter space is $\Theta = \left\{ \{\bar{N}(>S_{j,l})\}_j, \{\mathcal{D}_{S_{j,l}} \}_j, l^\circ, b^\circ \right\}$
{\color{responsecolor} i.e. for $n$ bins, there will be $2n+2$ parameters}.

\subsubsection{Monopole}
In a Bayesian framework, competing hypotheses can be ranked using Bayes factors. 
For completeness, we also define a null hypothesis by assuming that the observer is stationary, i.e. the observed source count in any pixel is uniform, or a monopole.
Analogous to the above discussion, we define the pixel-specific rate parameter for the flux bin to be
\begin{equation}
    \lambda_{i,j} = \bar{N}(>S_{j,l}) - \bar{N}(>S_{j,u})
\end{equation}
Now, the parameter space is $\Theta = \{\bar{N}(>S_{j,l})\}_j $, {\color{responsecolor} which gives $n$ parameters for $n$ bins}.

\subsubsection{Priors}
We adopt the following prior distributions for the free parameters of both the dipole and monopole models.
\begin{enumerate}
\item The number densities at the flux limits are sampled uniformly within a range of $\pm 5$ sources around the true number density at the corresponding flux limit: \\$\bar{N}(>S_{j,b}) \sim \mathcal{U}\left(\bar{N}_{\texttt{true}}(>S_{j,b})-5, \bar{N}_{\texttt{true}}(>S_{j,b})+5\right)$.
\item Dipole amplitudes are sampled uniformly between 0 and 0.1, i.e. $\mathcal{D}_S \sim \mathcal{U}(0,0.1)$
\item Directions are sampled uniformly over the surface of the sphere: $l^\circ \sim \mathcal{U}(0, 2\pi)$, and $b^\circ \sim \cos^{-1}(1 - 2u)$ for $u \sim \mathcal{U}(0,1)$.
\end{enumerate}
Here $\mathcal{U}$ denotes a uniform distribution within the mentioned limits.
{\color{responsecolor} In principle, we can use a broader prior range for both $\bar{N}(>S_{j,b})$ and  $\mathcal{D}_S$, but they only affect the absolute values of the Bayes factors and do not alter the qualitative outcomes of our results.}

\section{Results}\label{sec: results}
\subsection{Power-Law Distribution}\label{subsec:results_pareto}
\begin{figure*}[!ht]
\centering
\includegraphics[width=\linewidth]{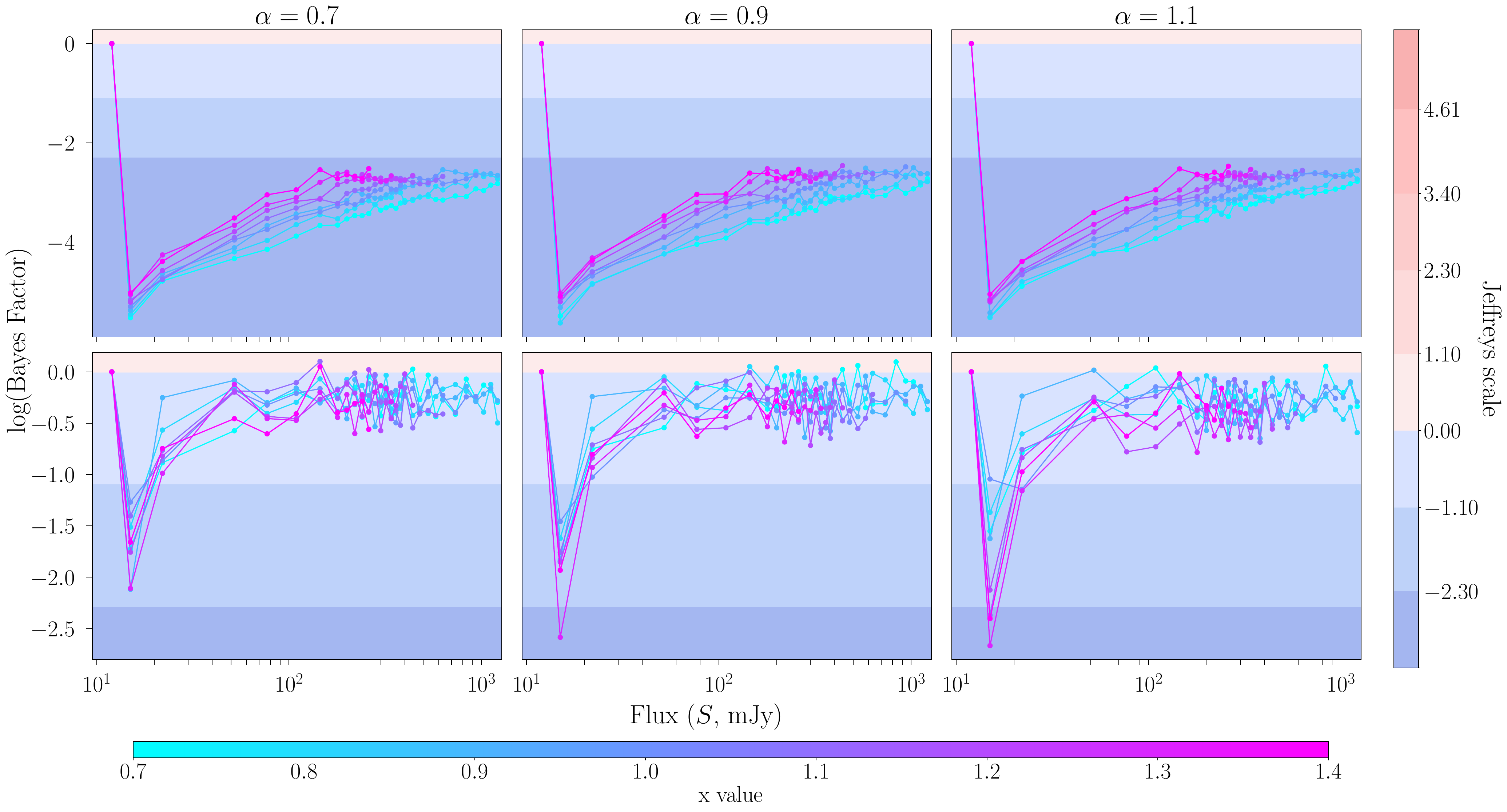}
\caption{\small Evolution of the Bayes factor (dipole vs monopole) with location of the flux cut for power-law LF. 
        The top panels show results for fitting a dipole amplitude at each flux limit, while the bottom panels show the results for fitting a single observer velocity $v$, shared across all flux limits.
        Each column corresponds to a different $\alpha$, and colours indicate the LF parameter $x$. 
        The left-most point in each curve represents the Bayes factor from the traditional E\&B analysis; the remaining points are obtained by bifurcating the catalogue at the indicated flux threshold and jointly analysing the two bins. 
        The vertical axis is shifted to show the difference relative to the traditional analysis.}
\label{fig:1}
\end{figure*}
The first LF we probe is the power-law distribution, which has been used in a number of important dipole studies to model the flux distribution of a source population.
This function maintains a constant slope of the survival function for any choice of limiting flux, i.e. $N(>S)\propto S^{-x}$, with $x = \texttt{constant}\, \forall S$.
Consequently, the dipole amplitude at all flux limits is constant provided that the source population has a common spectral index (which it is in this study).

As a starting point, we first probe the case where we bifurcate the catalogue into two flux bins by selecting an arbitrary flux cut.
Here and in the following joint analyses, we select the flux cuts by ensuring that they give at least 10 sources per pixel per flux bin, i.e. each flux bin should have at least $500,000$ sources.
To achieve this, we bifurcate the catalogue starting from a $15\, mJy$ flux cut and continue probing randomly selected flux cuts until the second flux bin has just fewer than $500,000$ sources.
This is done because our analysis shows that flux bins with sparsely populated pixels introduce shot noise, thereby affecting both the posterior and the Bayesian evidence.

We show the evolution of Bayes factors (dipole vs monopole) in the top panels of the Figure \ref{fig:1}.
Each panel differs in the spectral index of the sources, and the colour of the curves represents the value of the $x$ parameter used to generate the luminosity function.
In each curve, the left-most point represents the Bayes factors for the traditional E\&B. %analysis.
The Bayes factors at the other points on a curve were calculated by bifurcating the catalogue at the corresponding abscissa flux value, and then jointly analysing the two bins. 
Note that the Bayes factors at each point are consolidated over all the corresponding mock catalogues for the $(x,\alpha)$ combinations.
We shift the y-axis in each plot to show the relative difference between the traditional and the joint analyses.

We find that traditional E\&B provides a better description of the dipole parameters than joint analysis.
Curiously, if the flux cut for joint analysis is placed near the catalogue's limiting flux (where E\&B's fit is done), then the joint analysis is strongly disfavored with a (log) Bayes factor difference of $\approx 6$.
And as we shift the demarcation flux towards brighter values, we find that although the explanatory power of the joint analysis increases slightly, E\&B's approach %traditional analysis 
still provides the best grip on the matter dipole.
This seems to hint that, since both fits are performed on the same dataset, the joint analysis of a flux-binned catalogue is not a good approach to analyse the dataset.
To establish whether this is the case, we need to analyse the corresponding posterior distribution of the joint fits. 

{\color{responsecolor} As an example, we show the evolution of the posterior distributions (consolidated over the NS runs) of the fitted dipole parameters with the location of the flux cut, for $\alpha = 0.7$, $x=0.7$ in the Figure \ref{fig: A1}.
We find that the dipole direction, and dipole amplitude at lower flux limits are in excellent agreement with the input values, and have minimal (and stable level of) uncertanity regardless of the location of the flux cut.
But, the uncertanity in the dipole amplitude at the flux cut increases with the brightness of the flux cut.
The stability of dipole direction is understood by noting that these parameters are shared across both the flux bins, and hence, are constained by all the sources in the catalogue.
Similarly, the dipole amplitude at lower flux limit is constrained by the sources in the first flux bin, and since their count increases with shifting the flux cut to brighter values, the uncertanitity remains minimal, and stable.
But, the dipole amplitude at the flux cut is constrained by the sources in second flux bin, and since the count here decreases with shifting the flux cut, the uncertanity increases progressively.
Note that the corresponding posterior distributions for other choices of $\{\alpha, x\}$ show a similar trend.

To understand the Bayes factor curves, note that when the flux cut is placed in the vicinity of the lower flux limit, both the amplitudes are correlated with each other.
In this regime, we see a substantial decrease in support for the joint analysis.
If we move the flux cut towards brighter values, the amount of correlation decreases progressively, and this is augmented by the Bayes factors of both traditional and joint setup being almost equal to each other.
This suggests that the substantial decrease in support for joint analysis might stem from Occam's penalty, which penalizes overfitting.}
To address this, instead of fitting a dipole amplitude at each flux limit, we fit the observer's velocity $v$, which is shared across all flux limits.
This reduces the parameter space to $\Theta = \left\{ \{\bar{N}(>S_{j,b})\}_j, v, l^\circ, b^\circ \right\}$,
{\color{responsecolor} and fits $n+3$ free parameters for $n$ bins.}
We sample the velocity from a uniform distribution in the prior range 0 to 0.1, i.e. $v \sim  \mathcal{U}(0,0.1)$ \footnote{Note that a similar approach of rescaling dipole amplitudes to a shared velocity parameter has been used in \citep{Land-Strykowski:2025gkz}, to analyse different all-sky datasets jointly.}.

We repeat our analysis using this reduced parameter space, and show the evolution of Bayes factors in the bottom panels of Figure \ref{fig:1}. 
{\color{responsecolor} To demonstrate that we get the correct dipole parameters in the output, we show the evolution of the consolodated posterior distributions of $\{v,l^\circ, b^\circ\}$ for $\{\alpha, x\}=\{0.7, 0.7\}$ in Figure \ref{fig: A2}.
We find that all the parameters agree with the input values, are well constranined regadless of the location of the flux cut, and the uncertanity remains stable, which can be attributed to them being constrained by sources in both flux bins.}
This time, we find that the disagreement between the joint and traditional analyses reduces to an almost insignificant level.
We will address the decrease in Bayes factors for flux cuts near the lower flux limits in the discussion section.
This shows that, for a power-law LF, both E\&B and joint fit have equal explanatory power for the matter dipole - so, splitting the catalogue into disjoint flux bins offers no advantage for a pure power-law LF.

\subsection{Double Power-Law Distribution}
Next, we analyse the double power-law LF.
This function characterises a source population with a slightly steeper LF towards brighter fluxes.
We pin the transition flux limit at $200\ mJy$, because it enables us to place a sufficient number of flux cuts in either power-law region.
We first analyse the impact of dividing the catalogue into two bins, then discuss the impact of introducing a third flux bin.
\subsubsection{Two bins}
\begin{figure*}[!ht]
\centering
\includegraphics[width=\linewidth]{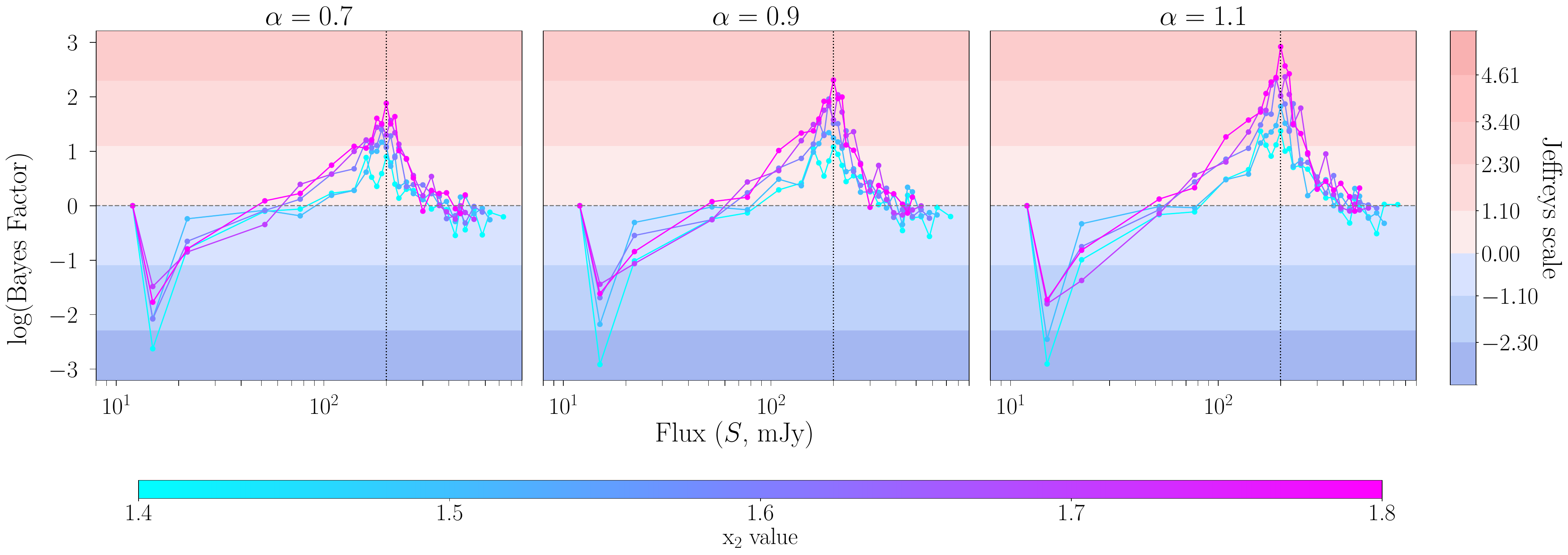}
\caption{\small As in the bottom panels of Figure \ref{fig:1}, where a shared observer's velocity $v$ is fitted across the flux limits, but for the double power-law LF.
        The colorbar indicates the magnitude of the LF parameter $x_2$.
        The vertical dashed line marks the location of the transition flux $S_{t}=200\ mJy$.}
\label{fig:2}
\end{figure*}
We perform the joint analysis of the two flux bins created by dividing the catalogue at a range of flux cuts by using the reduced parameter space $\Theta = \left\{ \{\bar{N}(>S_{j,b})\}_j, v, l^\circ, b^\circ \right\}$.
Figure \ref{fig:2} shows the evolution of the consolidated Bayes factors with the location of the flux cuts. 
Each subplot shows the result for a unique spectral index $\alpha$, while the colours represent the values for the second power law index $x_2$.
The first point in each curve gives the Bayes factors for the traditional E\&B, and the y-axis has been shifted to show the relative difference between this and the joint analysis for different flux cuts.
{\color{responsecolor} We find that the posterior distributions are in excellent agreement with the input values, and their uncertanity remains constant for any choice of flux cut.
We illustrate this by showing the evolution of the posterior of the shared parameters $\{v,l^\circ, b^\circ\}$ for $\{\alpha, x_2\}=\{0.7, 1.7\}$ in Figure \ref{fig: A3}.
}

Here, we find that, similar to the power-law LF, the Bayes factor curve initially dips, implying that the E\&B approach is preferred for low-magnitude flux cuts.
However, as we shift the flux cut towards higher values, we see that the Bayes factors initially increase, reach a maximum near the transition flux $S_{t}$ (marked with a vertical dotted line), and then decrease for higher flux cut values.
The increase depends directly on the magnitudes of $\alpha$ and $x_2$, with larger values giving stronger preference to joint analysis.
This shows that the joint analysis captures the transition between the two power-law regions and integrates this aspect of LF into the analysis, providing a better grip on the matter dipole.

\subsubsection{Three bins}
\begin{figure*}[!ht]
\centering
\begin{subfigure}{0.4\linewidth}
    \centering
    \includegraphics[width=\linewidth, trim={3cm 0cm 6cm 0cm},clip]{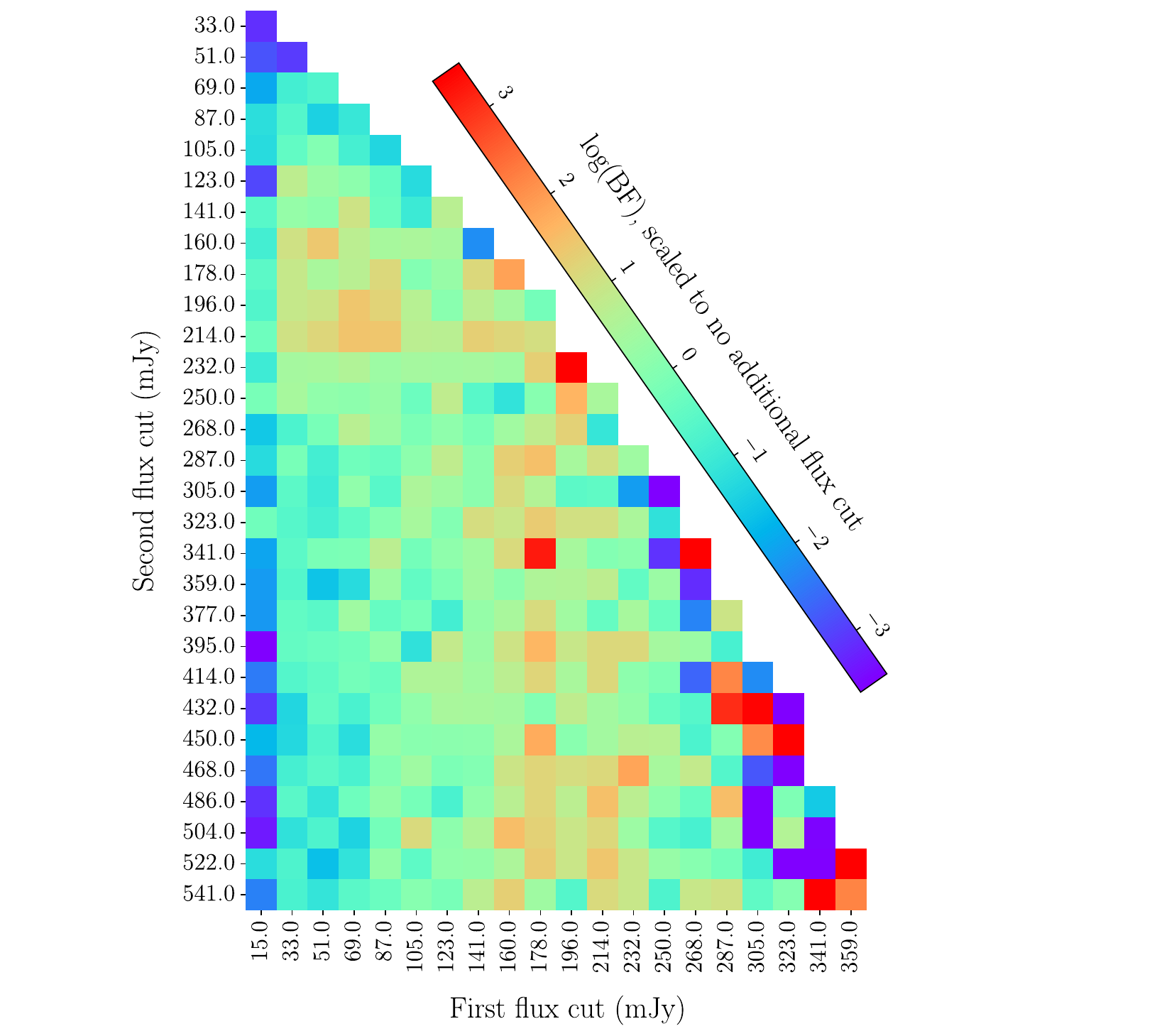}
    \caption{Double power-law LF}
\end{subfigure}
\hspace{0.05\linewidth}
\begin{subfigure}{0.47\linewidth}
    \centering
    \includegraphics[width=\linewidth, trim={3cm 0cm 3cm 0cm},clip]{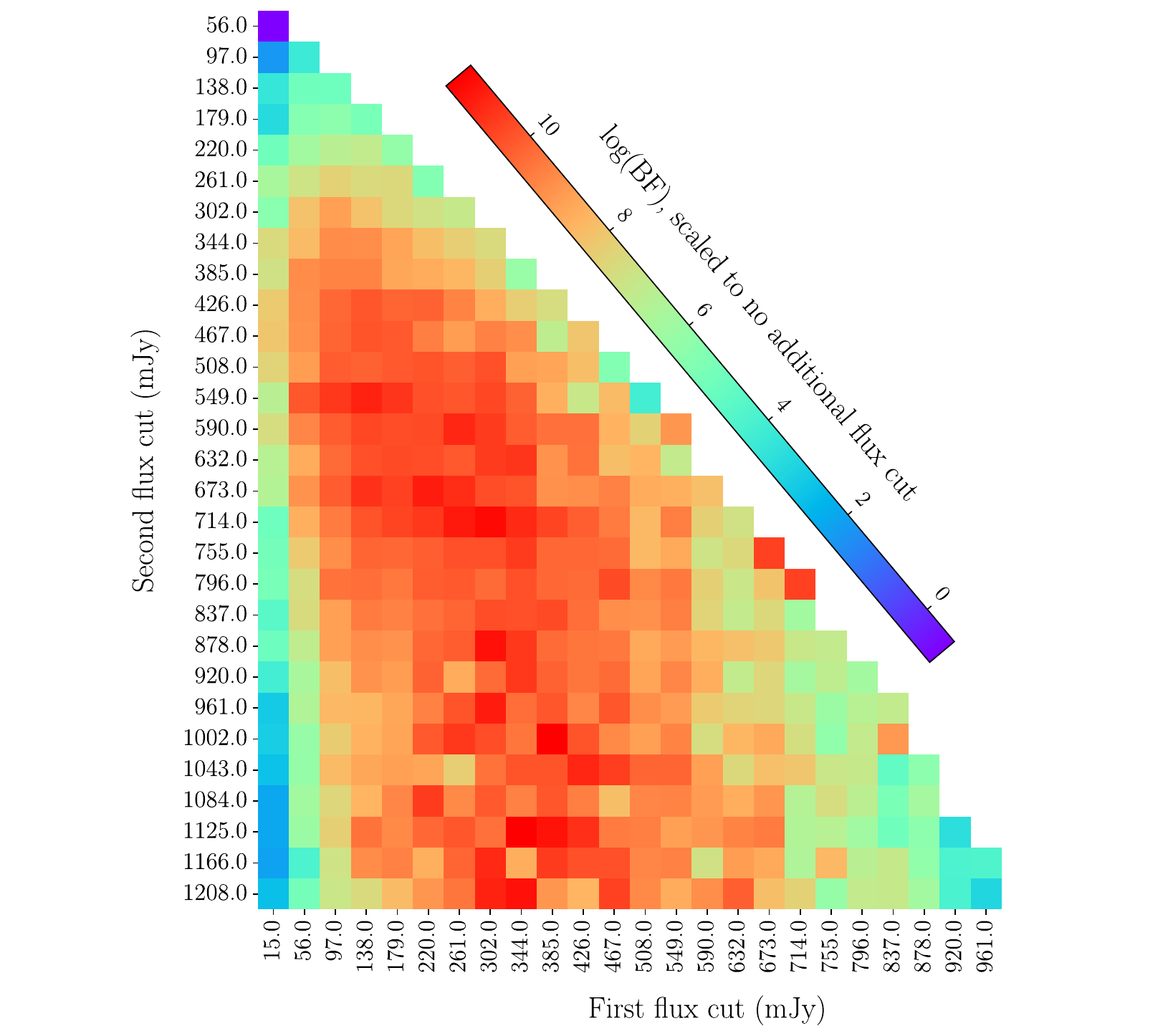}
    \caption{Schechter LF}
\end{subfigure}
\caption{\small Heatmap depicting the Bayes factors (dipole vs monopole) for three-bin joint analysis of mock catalogues constructed using a)- double power-law LF with parameters $\{ \alpha, x_2\}=\{0.9, 1.7\}$, b)- Schechter LF with parameters $\{\alpha, \upsilon\}=\{0.7, 0.7\}$. The colourbar has been shifted to show the difference relative to the traditional E\&B analysis.}
\label{fig:3}
\end{figure*}
\noindent The aforementioned shows that, for a double power-law LF, the two-bin joint analysis provides a better understanding of both the matter dipole and the flux distribution.
We now check if increasing the number of flux bins enhances this effect.
We use a range of flux cuts to divide the catalogue into three flux bins, and perform a joint analysis by using the reduced parameter set $\Theta = \left\{ \{\bar{N}(>S_{j,b})\}_j, v, l^\circ, b^\circ \right\}$.
We show the consolidated Bayes factors for joint analysis (relative to E\&B) for $\{\alpha, x_2\}=\{0.9, 1.7\}$ in the left panel of Figure \ref{fig:3}.
The Bayes factor for the traditional analysis is set to $0$ in the colourmap.
Note that the heatmap is not a perfect triangular plot, because, as we move towards the diagonal cells, the difference between the two flux cuts reduces, and we get $<10$ sources per pixel in the middle flux bin.
We only focus on the middle flux bin, because, the minimum and maximum values of the flux cut range, viz $15\, mJy$ and $541\, mJy$ are selected to give adequate source counts in the first and third flux bins.
In particular, the maxima was selected such that there were just over $500,000$ sources beyond it.
{\color{responsecolor} The recovered dipole parameters agree with inputs, and we show the posterior distribution for shared parameters $\{v,l^\circ, b^\circ\}$ for some combinations of the flux cuts in Figure \ref{fig: A4}.}
The heatmaps for other combinations of $\{\alpha, x_2\}$ differ quantitatively but qualitatively portray the same picture.

We find that the heatmap broadly shows four distinct regions: 
First, when the first flux cut is in the immediate vicinity of the catalogue's limiting flux, the E\&B is preferred over the joint fit.
Second, when the first flux cut lies well within the first power-law region and the second flux cut lies well within the second power-law region, we see that both methods have equal constraining power.
Third, when one of the two flux cuts is near $S_{t}$, the joint analysis yields slightly higher Bayes factors (within $ln2$) than the E\&B.
And fourth, when both flux cuts are well within the same power-law region, we see that E\&B performs significantly better than the joint analysis for most of the time.
If we compare the increase in Bayes factors near the transition flux in this scenario with its two-bin analysis counterpart, we find that the increase here is almost equal to the increase in the two-bin joint analysis.
This shows that adding an extra flux bin in analysis does not offer any additional insights into either the matter dipole or the flux distribution - rather, a two-bin joint fit with the flux cut near $S_{t}$ provides the strongest boost to the Bayesian evidence.

\subsection{Schechter Distribution}\label{subsec:results_schechter}
We now probe the Schechter LF, which has a power-law slope towards the fainter end and an exponential decrease at the brighter end.
This function is a popular choice for characterising the luminosity functions of a broad class of astronomical objects.
We pin the exponential decay limit at $S_{*}$ at $400\ mJy$, because it enables us to make a sizable number of adequately spaced flux cuts in both the power-law and exponential decay regions.
We first examine the impact of bifurcation, then discuss that of trifurcating the catalogue.
\subsubsection{Two bins}
\begin{figure*}[!ht]
\centering
\includegraphics[width=\linewidth]{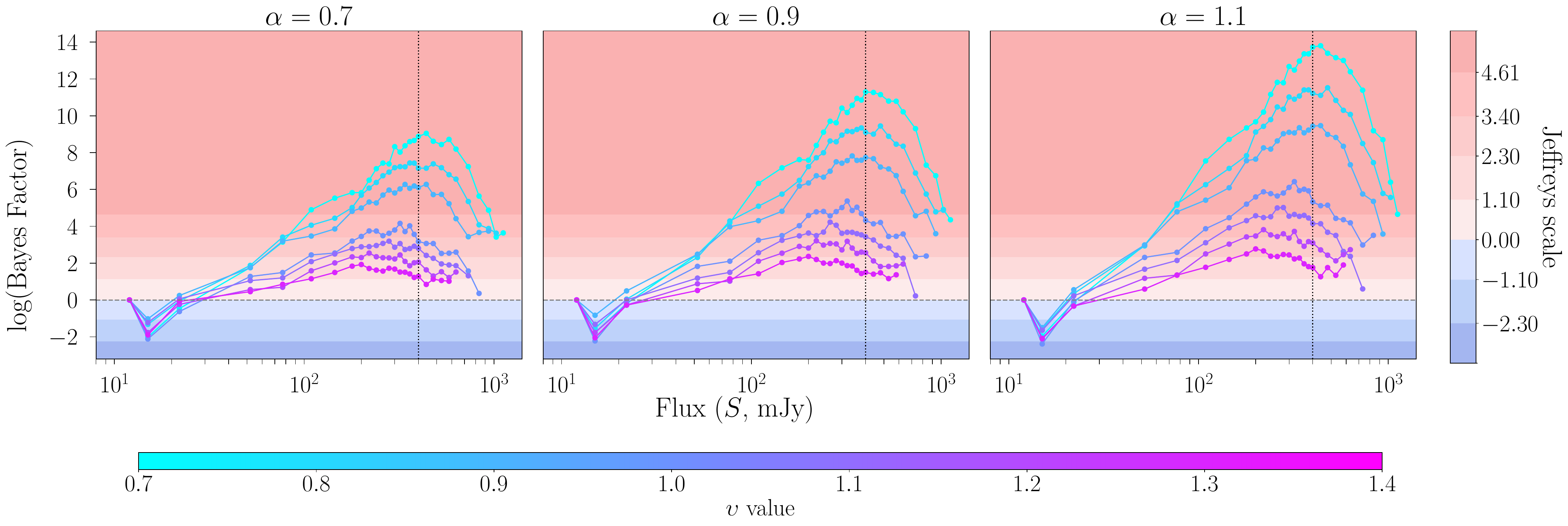}
\caption{\small As in Figure \ref{fig:2}, but for Schechter LF.
        The colourbar indicates the magnitude of the LF parameter $\upsilon$.
        The vertical dashed line marks the location of the limit of exponential decay $S_{*}=400\ mJy$.}
\label{fig:4}
\end{figure*}
\noindent We perform our joint analysis by bifurcating the catalogue using a range of flux cuts, and fit the reduced parameter space $\Theta = \left\{ \{\bar{N}(>S_{j,b})\}_j, v, l^\circ, b^\circ \right\}$.
We show the evolution of consolidated Bayes factors in Figure \ref{fig:4}. 
Here, each panel show the results for a particular value of {\color{responsecolor}$\alpha$}, while different colours represent different values of the parameter $\upsilon$.
As in the previous analyses, the first point on each curve represents the Bayes factors for the E\&B, and the y-axis is scaled to show the difference between this and joint analyses.
We again find that the recovered parameters agree with our input parameters.
{\color{responsecolor} As an example, we show the evolution of posterior distribution for the shared parameters $\{v,l^\circ, b^\circ\}$ for $\{\alpha, \upsilon\}=\{0.7, 0.9\}$ in Figure \ref{fig: A5}, where the parameters remain well constrained and their uncertanties remain stable.}

This time, we find that even though support for joint analysis decreases initially, as we move the flux cut towards the brighter region of the luminosity function, the Bayes factors first increase overwhelmingly, then decrease significantly as the flux cut approaches the upper flux limit.
We also observe that the increase in Bayes factors is more pronounced for lower values of $\upsilon$ as compared to that for higher $\upsilon$.
This means that the joint analysis performs better for an LF with a shallower power-law region and a more pronounced exponential decay.
Additionally, as in the double power-law LF, we observe that the increase is directly proportional to $\alpha$ - for the same $\upsilon$, a larger $\alpha$ gives a higher peak.

\subsubsection{Three bins}
We perform our joint analysis by trifurcating the catalogue into three flux bins for a range of flux cuts, using the reduced parameter space $\Theta = \left\{ \{\bar{N}(>S_{j,b})\}_j, v, l^\circ, b^\circ \right\}$.
In the right panel of Figure \ref{fig:3}, we show the Bayes factors for the joint analysis for $\{\alpha, \upsilon\}=\{0.7, 0.7\}$.
Similar to the double power-law analysis, the colorbar is scaled so that the Bayes factor for the traditional analysis corresponds to the $0$ point of the colour scheme.
The heatmaps for other combinations of $\{\alpha, \upsilon\}$ are qualitatively similar, but differ in their exact range of Bayes factors.
Similar to previous cases, we find that the recovered parameters agree with our input parameters.
{\color{responsecolor} The recovered dipole parameters agree with inputs, and we show the posterior distribution for shared parameters $\{v,l^\circ, b^\circ\}$ for some combinations of the flux cuts in Figure \ref{fig: A6}.
Similar to the previously discussed cases, the uncertanities remain constant, because the shared parameters are being constrained by sources in all the flux bins.}

Here, we find that if the first flux cut is placed quite near the lower flux limit of the catalogue, then, if the second flux cut is not in the middle of the LF, there is almost equal support for both methods.
While if the second flux cut is in the middle of the LF, there is an overwhelming preference for the joint analysis with a maximum $\Delta log(BF) \approx 8$.
As we shift the first flux cut away from the lower limit and the second flux cut is a bit further towards the brighter fluxes, then we see that joint analyses provide a better grip on the dipole parameters than the usual E\&B.
In fact, the highest support stems from placing both cuts somewhere in the middle of the LF, away from both the upper and lower limits of the catalogue, and ensuring that there is a sufficient space between them.
Finally, if both the flux cuts are either well in the power-law region (towards the upper corner of the heatmap) or well in the exponential decay region (towards the bottom right corner), then although the support is somewhat higher than E\&B, it is still lower than the case where both flux cuts are in the middle region of the LF.

\section{Discussion \& Conclusion}\label{sec: discussion}
\subsection{Shape of the Bayes factor curves}
\noindent In the previous section, for any $(LF, \alpha)$ combination, we observed a significant shift in Bayes factors as we switched from traditional to joint analysis.
We saw two major trends: first, if we placed the flux cut (or one of the flux cuts) near the catalogue's limiting flux, support for the joint analysis decreased.
Second, for non-power-law LFs, as we shifted the flux cut(s) towards brighter fluxes, the joint analysis was decisively favoured.

\begin{figure*}[!t]
\centering
\includegraphics[width=\linewidth]{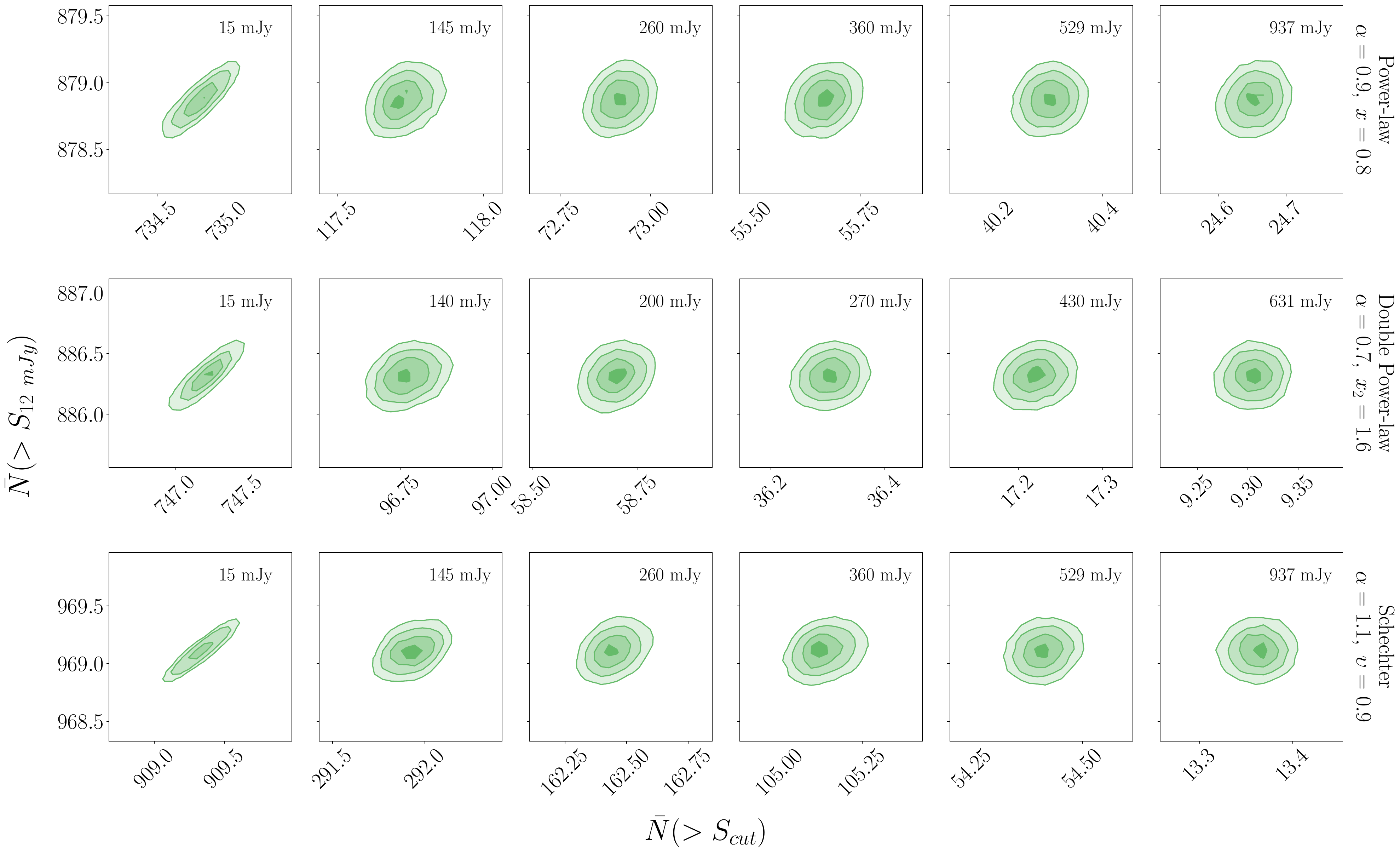}
\caption{\small 
        Evolution of the $\{\bar{N}(>S_{12\, mJy}), \bar{N}(>S_{cut})\}$ 2-D histogram of the consolidated Posterior distributions for the two-bin joint analysis of some of the LFs.
        Each row showcases the 2-D histograms for a range of increasing flux cuts (mentioned in the subplot's inset) for the LF mentioned in the right label of each row.
        The contours enclose $12\%$, $39\%$, $68\%$ and $86\%$ of the distribution.}
\label{fig:5}
\end{figure*}
\begin{figure*}[!ht]
\centering
\includegraphics[width=0.5\linewidth]{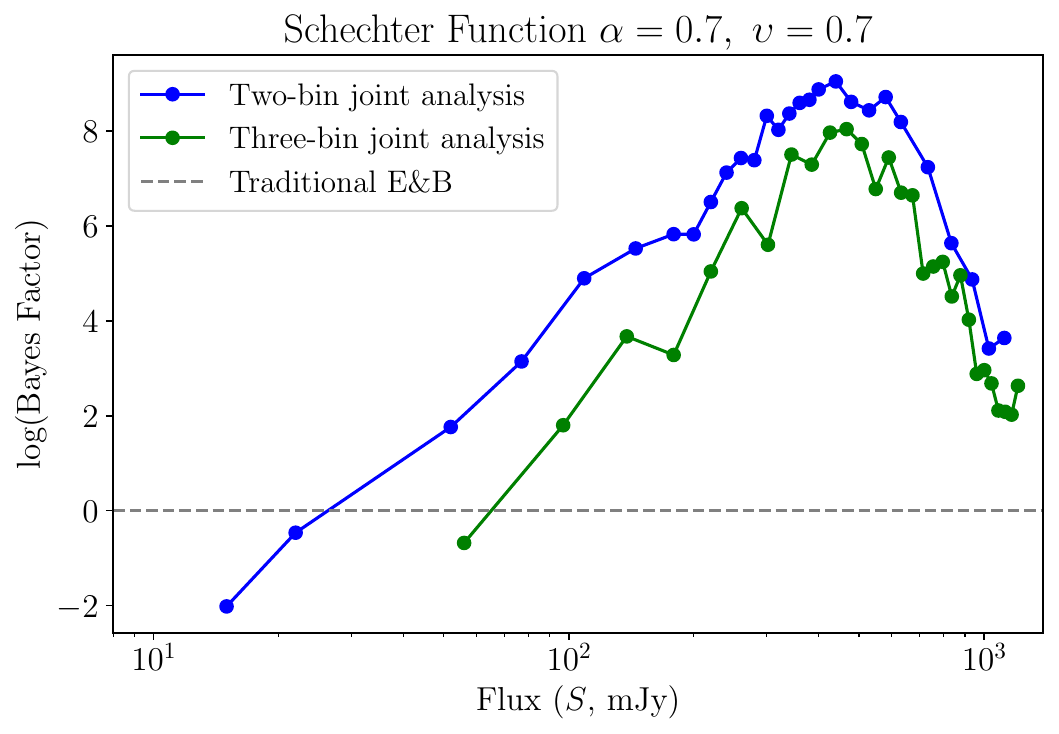}
\caption{\small 
        Comparison of the evolution of consolidated Bayes factors for two-bin and three-bin joint analysis of the Schechter LF with parameters $\{\alpha,\upsilon\} = \{0.7,0.7\}$.
        For the two-bin curve, the x-axis marks the location of the flux cut, while it marks the location of the second flux cut for the three-bin curve.
        The y-axis for both curves is shifted to show the relative difference between the joint analyses and the E\&B analysis, with the latter fixed at 0.
        }
\label{fig:6}
\end{figure*}

We first examine the decrease in support for joint analysis over E\&B if we place the flux cut (or one of the flux cuts) near the catalogue's limiting flux. 
This trend is observed in almost all of the above-discussed cases, except for the three bin joint analysis of Schechter LF, which we will discuss shortly. 
The generality rules out its relation to any LF-related systematic; rather, it hints towards its statistical origins.
This dip is similar to the dip in Bayes factors in Section \ref{subsec:results_pareto}, which was generated due to a correlation between $\{\mathcal{D}_{S_{12mJy}}, \mathcal{D}_{S_{\texttt{cut}}}\}$.
There, the dip became less significant once we reduced the parameter space by rescaling the amplitudes in terms of the observer's velocity.
Taking a cue, if we again look at the parameter space, $\Theta = \left\{ \{\bar{N}(>S_{j,b})\}_j, v, l^\circ, b^\circ \right\}$, we infer that the decrease here might be caused by a correlation between the number density parameters $\{\bar{N}(>S_{j,b})\}_j$.
To check if this is so, we plot the 2D histograms between the posterior samples of $\bar{N}(>S_{12\ mJy})$ and $\bar{N}(>S_{cut})$ for a range of flux cuts for the two bin joint analysis of all three luminosity functions in Figures \ref{fig:5}.
We see that for $S_{cut}=15\ mJy$, both $\bar{N}(>S_{12\ mJy})$ and $\bar{N}(>S_{cut})$ are highly correlated with each other, while the correlation is negligible for other flux cuts.
The correlation implies that $\bar{N}(>S_{15\ mJy})=k\bar{N}(>S_{12\ mJy})$ (where $k$ is a scaling constant), i.e. we are using more than the required number of parameters.
This activates Occam's penalty, which decreases the Bayes factors when $S_{cut}$ is near $12\ mJy$.

Coming to the three bin joint analysis of Schechter LF, in Figure \ref{fig:3}, we observe that if the first flux cut is placed at $15\ mJy$, then the joint analysis is favoured for almost all choices of the second cut.
This seems to contradict Occam's penalty explanation for the decrease in support for joint analysis.
To explain this scenario, in Figure \ref{fig:6}, we present a comparison between the $15\ mJy$ column of Figure \ref{fig:3}, with the corresponding two-bin curve of Figure \ref{fig:4}.
We find that both curves align well and have peaks in similar regions.
But the Bayes factors for the three-bin analysis are always less than those of the two-bin analysis, i.e., the Bayes factor decreases upon introducing an additional flux cut near the limiting flux.
Thus, even though the correlation between $\bar{N}(>S_{12\ mJy})$ and $\bar{N}(>S_{15\ mJy})$ injects an Occam's penalty, the magnitude of the penalty is not sufficient to phase out the increase in Bayes factors due to the shape of the luminosity function.

Next, we examine the increase in Bayes factors for the joint analyses of the double power-law and Schechter LFs.
We first note that for a double power-law LF, the magnitude of the increase in Bayes factors in two-bin analysis depends directly on the difference between the magnitudes of $x_1$ and $x_2$, with a larger difference yielding a higher peak in the curves.
As the difference between $x_1$ and $x_2$ decreases and the LF approaches a pure power-law curve, the peak becomes negligible, and we get equal support for both traditional and joint analysis. 
In its three-bin analysis, we do not observe any significant increase in support for dipole over the two-bin case.
In fact, only when one of the two flux cuts is placed near the transition region, the three bin analysis gives comparable Bayes factors to the two bin counterpart.
In all other cases, it performs at par with the traditional analysis.

For the Schechter distribution, the most significant increase in Bayes factors for the two bin analysis occurs when we use a low $\upsilon$, which gives a pronounced transition to the exponential decay region.
As we increase $\upsilon$, and the LF starts to approach a pure power-law, the increase in Bayes factors becomes negligible.
Similarly, for its three-bin analysis, we only get a decisive verdict in favour of joint analysis when both flux cuts are placed in the non-power-law region, where the source counts drop significantly upon a small increment in flux.
Taken together, these features show that if the LF has pronounced deviations from a power-law structure, then placing a flux cut in the deviating region gives a better grip on the matter dipole.
In other words, capturing the global shape of the LF through joint analysis of flux-binned catalogue gives a stronger grasp of the underlying matter dipole signal as compared to traditional E\&B, which only analyses the local shape of the LF at the limiting flux.

\subsection{How many sources are needed?}
\noindent Now that we understand the origin of the increase in Bayes factors for the joint analysis, the next question concerns the minimum number of sources required to robustly apply this method.
While our results establish that joint analysis provides a better characterisation of the matter dipole, its practical recovery is fundamentally limited by survey characteristics and the total number of observed sources. 
Determining this minimum source-count threshold is therefore essential for connecting theoretical expectations with observational feasibility. 
To address this, we examine how the Bayes factors evolve as the number of sources is progressively reduced.

\begin{figure*}[t]
\centering
\includegraphics[width=\linewidth]{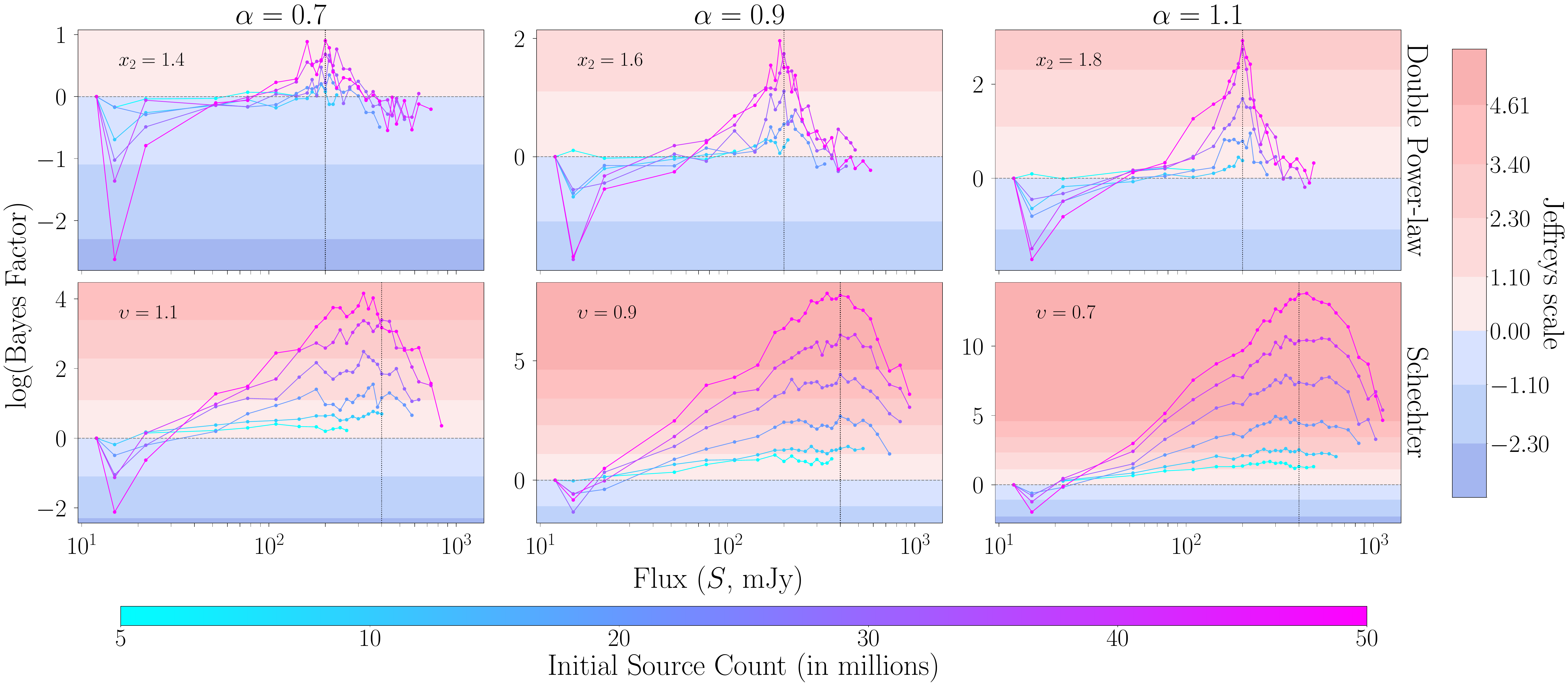}
\caption{\small 
        Evolution of the Bayes factors (dipole vs monopole) with a change in initial source counts for some of the LFs studied in this work.
        While each column corresponds to a different $\alpha$, each row corresponds to a different LF type with the corresponding free parameter mentioned in the inset of each plot.
        The colours represent the initial source count used to make the mock catalogues.
        Other properties of the plot are similar to the Figure \ref{fig:2}.
        }
\label{fig:7}
\end{figure*}

In Figure \ref{fig:7}, we show the evolution of Bayes factor curves for two bin joint analyses of some randomly selected $(LF,\alpha)$ with the initial source counts used to construct the mocks.
We observe the following salient features in each subplot.
Firstly, for an initial source count of $\approx$ 10 million sources, even though the Bayes factor curves reach the regions beyond the transition fluxes of $200\ mJy$ and $400 \ mJy$ for double power-law and Schechter LFs, respectively, there is only a marginal difference between the Bayes factors for traditional and joint analysis.
This trend holds regardless of the location of the flux cut and suggests that merely having a decent number of sources in the non-power-law regime does not give any significant improvement in the analysis's outcome.
For high spectral indices ($\alpha = 0.9, 1.1$), if the initial source count is increased beyond 20 million, we find that joint analysis is decisively favoured over the traditional analysis for both LFs.
Whereas for a low spectral index of $0.7$, we need around 30 million initial sources to get decisive support for the joint analysis.

One important point deserves clarification here. 
The initial source count of 20 million (for $\alpha = 0.9,1.1$ or 30 million for $\alpha=0.7$)  is for constructing the isotropic sky map, with a limiting flux of $10\ mJy$.
However, as we have mentioned in Section \ref{sec: simulated skies}, after we inject a dipole in this map, we select all sources above $12\ mJy$, as this removes the distorted region of the LF, which occurred due to Doppler boosting.
Therefore, the actual source counts in the catalogue used for analysis are a bit less than 20 million.
After accounting for this trimming, we find that having at least 18 (or 27) million sources in the final catalogue will provide sufficient sources for this analysis.

\subsection{Conclusions \& Observational Feasibility}
In this study, we discussed a new method of calculating the matter dipole from all-sky catalogues of cosmological sources.
Our method builds on the classic Ellis \& Baldwin test and integrates the flux of sources in the analysis.
We calculate the matter dipole by simultaneously calculating the matter dipole at a number of flux limits.
We achieve this by first dividing the catalogue into multiple disjoint flux bins and then jointly analysing all the flux bins to calculate the matter dipole at their boundaries.
In the process, we reparameterize the dipole amplitudes at each flux limit/cut in terms of the observer's velocity, which, along with the dipole direction, is shared across all flux bins.
Additionally, we fit the cumulative source count at the limiting flux and the flux cut.
We calculate our Bayes factors by selecting the monopole as the reference hypothesis. 

We find that if the source population's LF follows a pure power-law distribution, our method offers no significant advantage over E\&B's method.
But if the source population follows another LF, such as the double-power-law or Schechter function, then our method provides improved grip over the matter dipole parameters.
The level of improvement depends on four factors: the location of the bin boundaries, the deviation of LF from a power law, the SED of the population, and the number of sources in the catalogue.
If we place the bin boundary close to the catalogue's limiting flux, support for our method decreases relative to the E\&B.
This occurs because of the strong correlation between the cumulative number counts for the limiting flux and the bin boundary, which activates Occam's penalty, and decreases the Bayes factors by $\approx ln 2$.
On the other hand, if we place the flux cut in the region where the LF deviates from a power-law structure, we find that the Bayes factors for joint analysis increase.
A larger deviation from the power-law LF gives a larger increase in Bayes factors, and enables an increase in the number of flux cuts.
As an example of this, a double-power-law LF gives maximum grip on matter dipole if we place a single flux cut in the proximity of the transition flux, while a Schechter LF allows for more flux cuts in the exponential decay region.
The increase in Bayes factors upon joint analysis also depends on the spectral index $\alpha$ - a source population with a higher $\alpha$ gives a larger increase in Bayes factors.
Finally, a larger source count yields more sources on either side of a flux cut placed in the non-power-law region, leading to a higher Bayes factor.

Having established the requirements for obtaining a robust outcome using joint analysis, we turn to the question of observational feasibility. 
We now discuss how ongoing and planned surveys can be used to detect the dipole using the method derived in our simulations.
This analysis allows us to assess which surveys can provide meaningful constraints and identify observational regimes where a significant detection is likely.
Below, we discuss the feasibility of our methods for two regimes in which the discrepancy between the kinematic and matter dipoles has posed a significant challenge for the CP.

For \emph{radio} wavelengths, flagship simulations like T-RECS \citep{Bonaldi:2018xfm} predict a non-power-law LF for upcoming radio continuum surveys. 
As an example, in T-RECS II \citep[Figure 3 of ][]{Bonaldi:2023fnb}, where the LFs have been calculated at $1.4\ GHz$ after accounting for H-I emission, we find that for the 1-1.2 redshift bin, the LF shows a characteristic non-power-law shape for Luminosities which are in the range $10^{23}-10^{26}\ W/Hz$, or equivalently flux range of $3.9-3900\ \mu Jy$ (assuming $\Lambda$-CDM cosmology and $\alpha=0.8$ to account for k-correction).
Multiband simulations like SPRITZ \citep[Figure 22 of][]{Bisigello_2021} also predict a characteristic non-power-law LF in a similar observational regime.
This suggests that the redshift cumulated LF will be characteristically non-power-law for $1.4 GHz$ in the $\mu Jy$ to sub-$mJy$ regime. 
Ongoing surveys such as the Evolutionary Map of the Universe (EMU) \citep{emu, emu-casda} and upcoming surveys such as the Square Kilometre Array (SKA)\footnote{\url{https://www.skao.int/en}} will probe the sky in this band and flux ranges.
EMU expects to observe 40 million sources in the southern sky at sub-$mJy$ fluxes, which extrapolates to a high all-sky source count and would provide an excellent dataset for joint analysis \citep{camera2015}.
Similarly, SKA-low expects to observe $9\times10^8$ galaxies over $30,000\ deg^2$, up to $z\approx2$ with $\mu Jy$ sensitivities, providing sufficient information for joint analysis of the flux-binned catalogue.

 In the \emph{infrared} domain, E\&B fit is done on quasar candidates. 
Results from the Sptizer telescope have shown that the quasar candidates follow a broken/double power-law LF \citep[see Figure 4 of ][]{Lacy:2015tha}.
This suggests that a sizable population of quasar candidates should allow for a flux-binned analysis.
Euclid \citep{Euclid2025} will observe around 8.1 million quasar candidates over $14,000\ deg^{2}$ of the extragalactic sky, which translates to around 24 million all-sky sources.
Since quasars have a high $\alpha \sim 1-2$ in IR, this source count should be sufficient to perform a flux-binned analysis of the source population.

Building on the framework presented above, future work can extend this analysis in several directions. 
We outline a few of them here.
At present, flux-cut locations are selected randomly; developing a formulation to determine these locations from first principles for a general luminosity function would improve the speed and efficiency of the method. 
Similarly, identifying the optimal number of flux cuts without resorting to simulations would further simplify the analysis. 
Another line of work involves reducing the number of sources required by this method, thereby broadening its applicability to existing datasets.
Finally, using this method to analyse the higher-order multipoles in the number density contrast can enhance our understanding of real survey catalogues.
In conclusion, while this method provides a better understanding of the matter dipole, further refinements will enhance its usability.
This will be the topic of a future publication.

\acknowledgments
VM acknowledges financial support from the Physics Foundation Scholarship.
GFL thanks Chris Clarkson and Charles Dalang for hosting him at the QMUL and for the discussions that sowed the seed of this study.
{\color{responsecolor} We thank the referee for their comments.
This work used the School of Physics, University of Sydney's HPC cluster.}
This work made use of the \textsc{python} packages
\textsc{UltraNest} \citep{skilling2004, skilling2006, 2021JOSS....6.3001B}, \textsc{Jax} \citep{jax2018github}, \textsc{healpy} \citep{Gorski2005,Zonca2019}, \textsc{numpy} \citep{harris2020}, \textsc{matplotlib} \citep{hunter2007}, \textsc{scipy} \citep{scipy2020}, \textsc{pandas}\citep{reback2020} and \textsc{astropy} \citep{astropy2022}.

\bibliographystyle{JHEP}
\bibliography{main.bib}

@article{kass1995,
  issn      = {01621459},
  url       = {http://www.jstor.org/stable/2291091},
  author    = {Robert E. Kass and Adrian E. Raftery},
  journal   = {Journal of the American Statistical Association},
  number    = {430},
  pages     = {773--795},
  publisher = {[American Statistical Association, Taylor & Francis, Ltd.]},
  title     = {Bayes Factors},
  volume    = {90},
  year      = {1995}
}

@inproceedings{skilling2004,
  author    = {{Skilling}, John},
  title     = {{Nested Sampling}},
  booktitle = {Bayesian Inference and Maximum Entropy Methods in Science and Engineering: 24th International Workshop on Bayesian Inference and Maximum Entropy Methods in Science and Engineering},
  year      = 2004,
  editor    = {{Fischer}, Rainer and {Preuss}, Roland and {Toussaint}, Udo Von},
  series    = {American Institute of Physics Conference Series},
  volume    = {735},
  month     = nov,
  pages     = {395-405},
  doi       = {10.1063/1.1835238},
  adsurl    = {https://ui.adsabs.harvard.edu/abs/2004AIPC..735..395S}
}

@article{skilling2006,
  author    = {{Skilling}, John},
  title     = {{Nested sampling for general Bayesian computation}},
  volume    = {1},
  journal   = {Bayesian Analysis},
  number    = {4},
  publisher = {International Society for Bayesian Analysis},
  pages     = {833 -- 859},
  year      = {2006},
  doi       = {10.1214/06-BA127},
  url       = {https://doi.org/10.1214/06-BA127}
}

@ARTICLE{2016S&C....26..383B,
       author = {{Buchner}, Johannes},
        title = "{A statistical test for Nested Sampling algorithms}",
      journal = {Statistics and Computing},
         year = 2016,
        month = jan,
       volume = {26},
       number = {1-2},
        pages = {383-392},
          doi = {10.1007/s11222-014-9512-y},
archivePrefix = {arXiv},
       eprint = {1407.5459},
 primaryClass = {stat.CO},
       adsurl = {https://ui.adsabs.harvard.edu/abs/2016S&C....26..383B}
}

@ARTICLE{2019PASP..131j8005B,
       author = {{Buchner}, Johannes},
        title = "{Collaborative Nested Sampling: Big Data versus Complex Physical Models}",
      journal = {\pasp},
         year = 2019,
        month = oct,
       volume = {131},
       number = {1004},
        pages = {108005},
          doi = {10.1088/1538-3873/aae7fc},
archivePrefix = {arXiv},
       eprint = {1707.04476},
 primaryClass = {stat.CO},
       adsurl = {https://ui.adsabs.harvard.edu/abs/2019PASP..131j8005B}
}

@article{Gorski2005,
  author   = {{G{\'o}rski}, K.~M. and {Hivon}, E. and {Banday}, A.~J. and 
              {Wandelt}, B.~D. and {Hansen}, F.~K. and {Reinecke}, M. and 
              {Bartelmann}, M.},
  title    = {{HEALPix: A Framework for High-Resolution Discretization and Fast Analysis of Data Distributed on the Sphere}},
  journal  = {\apj},
  eprint   = {arXiv:astro-ph/0409513},
  year     = 2005,
  month    = apr,
  volume   = 622,
  pages    = {759-771},
  doi      = {10.1086/427976},
  adsurl   = {http://adsabs.harvard.edu/abs/2005ApJ...622..759G}
}

@article{hunter2007,
  author    = {Hunter, J. D.},
  title     = {Matplotlib: A 2D graphics environment},
  journal   = {Computing in Science \& Engineering},
  volume    = {9},
  number    = {3},
  pages     = {90--95},
  publisher = {IEEE COMPUTER SOC},
  doi       = {10.1109/MCSE.2007.55},
  year      = 2007
}

@article{Zonca2019,
  doi       = {10.21105/joss.01298},
  url       = {https://doi.org/10.21105/joss.01298},
  year      = {2019},
  month     = mar,
  publisher = {The Open Journal},
  volume    = {4},
  number    = {35},
  pages     = {1298},
  author    = {Andrea Zonca and Leo Singer and Daniel Lenz and Martin Reinecke and Cyrille Rosset and Eric Hivon and Krzysztof Gorski},
  title     = {healpy: equal area pixelization and spherical harmonics transforms for data on the sphere in Python},
  journal   = {Journal of Open Source Software}
}

@misc{reback2020,
  author    = {{The Pandas Development Team}},
  title     = {pandas-dev/pandas: Pandas},
  month     = feb,
  year      = 2020,
  publisher = {Zenodo},
  version   = {latest},
  doi       = {10.5281/zenodo.3509134},
  url       = {https://doi.org/10.5281/zenodo.3509134}
}

@article{harris2020,
  title     = {Array programming with {NumPy}},
  author    = {Charles R. Harris and K. Jarrod Millman and St{\'{e}}fan J.
               van der Walt and Ralf Gommers and Pauli Virtanen and David
               Cournapeau and Eric Wieser and Julian Taylor and Sebastian
               Berg and Nathaniel J. Smith and Robert Kern and Matti Picus
               and Stephan Hoyer and Marten H. van Kerkwijk and Matthew
               Brett and Allan Haldane and Jaime Fern{\'{a}}ndez del
               R{\'{i}}o and Mark Wiebe and Pearu Peterson and Pierre
               G{\'{e}}rard-Marchant and Kevin Sheppard and Tyler Reddy and
               Warren Weckesser and Hameer Abbasi and Christoph Gohlke and
               Travis E. Oliphant},
  year      = {2020},
  month     = sep,
  journal   = {Nature},
  volume    = {585},
  number    = {7825},
  pages     = {357--362},
  doi       = {10.1038/s41586-020-2649-2},
  publisher = {Springer Science and Business Media {LLC}},
  url       = {https://doi.org/10.1038/s41586-020-2649-2}
  }

@article{scipy2020,
  author  = {Virtanen, Pauli and Gommers, Ralf and Oliphant, Travis E. and
             Haberland, Matt and Reddy, Tyler and Cournapeau, David and
             Burovski, Evgeni and Peterson, Pearu and Weckesser, Warren and
             Bright, Jonathan and {van der Walt}, St{\'e}fan J. and
             Brett, Matthew and Wilson, Joshua and Millman, K. Jarrod and
             Mayorov, Nikolay and Nelson, Andrew R. J. and Jones, Eric and
             Kern, Robert and Larson, Eric and Carey, C J and
             Polat, {\.I}lhan and Feng, Yu and Moore, Eric W. and
             {VanderPlas}, Jake and Laxalde, Denis and Perktold, Josef and
             Cimrman, Robert and Henriksen, Ian and Quintero, E. A. and
             Harris, Charles R. and Archibald, Anne M. and
             Ribeiro, Ant{\^o}nio H. and Pedregosa, Fabian and
             {van Mulbregt}, Paul and {SciPy 1.0 Contributors}},
  title   = {{{SciPy} 1.0: Fundamental Algorithms for Scientific
             Computing in Python}},
  journal = {Nature Methods},
  year    = {2020},
  volume  = {17},
  pages   = {261--272},
  adsurl  = {https://rdcu.be/b08Wh},
  doi     = {10.1038/s41592-019-0686-2}
}

@ARTICLE{2021JOSS....6.3001B,
       author = {{Buchner}, Johannes},
        title = "{UltraNest - a robust, general purpose Bayesian inference engine}",
      journal = {The Journal of Open Source Software},
         year = 2021,
        month = apr,
       volume = {6},
       number = {60},
          eid = {3001},
        pages = {3001},
          doi = {10.21105/joss.03001},
archivePrefix = {arXiv},
       eprint = {2101.09604},
 primaryClass = {stat.CO},
       adsurl = {https://ui.adsabs.harvard.edu/abs/2021JOSS....6.3001B}
}

@ARTICLE{astropy2022,
       author = {{Astropy Collaboration} and {Price-Whelan}, Adrian M. and {Lim}, Pey Lian and {Earl}, Nicholas and {Starkman}, Nathaniel and {Bradley}, Larry and {Shupe}, David L. and {Patil}, Aarya A. and {Corrales}, Lia and {Brasseur}, C.~E. and {N{"o}the}, Maximilian and {Donath}, Axel and {Tollerud}, Erik and {Morris}, Brett M. and {Ginsburg}, Adam and {Vaher}, Eero and {Weaver}, Benjamin A. and {Tocknell}, James and {Jamieson}, William and {van Kerkwijk}, Marten H. and {Robitaille}, Thomas P. and {Merry}, Bruce and {Bachetti}, Matteo and {G{"u}nther}, H. Moritz and {Aldcroft}, Thomas L. and {Alvarado-Montes}, Jaime A. and {Archibald}, Anne M. and {B{'o}di}, Attila and {Bapat}, Shreyas and {Barentsen}, Geert and {Baz{'a}n}, Juanjo and {Biswas}, Manish and {Boquien}, M{'e}d{'e}ric and {Burke}, D.~J. and {Cara}, Daria and {Cara}, Mihai and {Conroy}, Kyle E. and {Conseil}, Simon and {Craig}, Matthew W. and {Cross}, Robert M. and {Cruz}, Kelle L. and {D'Eugenio}, Francesco and {Dencheva}, Nadia and {Devillepoix}, Hadrien A.~R. and {Dietrich}, J{"o}rg P. and {Eigenbrot}, Arthur Davis and {Erben}, Thomas and {Ferreira}, Leonardo and {Foreman-Mackey}, Daniel and {Fox}, Ryan and {Freij}, Nabil and {Garg}, Suyog and {Geda}, Robel and {Glattly}, Lauren and {Gondhalekar}, Yash and {Gordon}, Karl D. and {Grant}, David and {Greenfield}, Perry and {Groener}, Austen M. and {Guest}, Steve and {Gurovich}, Sebastian and {Handberg}, Rasmus and {Hart}, Akeem and {Hatfield-Dodds}, Zac and {Homeier}, Derek and {Hosseinzadeh}, Griffin and {Jenness}, Tim and {Jones}, Craig K. and {Joseph}, Prajwel and {Kalmbach}, J. Bryce and {Karamehmetoglu}, Emir and {Ka{l}uszy{'n}ski}, Miko{l}aj and {Kelley}, Michael S.~P. and {Kern}, Nicholas and {Kerzendorf}, Wolfgang E. and {Koch}, Eric W. and {Kulumani}, Shankar and {Lee}, Antony and {Ly}, Chun and {Ma}, Zhiyuan and {MacBride}, Conor and {Maljaars}, Jakob M. and {Muna}, Demitri and {Murphy}, N.~A. and {Norman}, Henrik and {O'Steen}, Richard and {Oman}, Kyle A. and {Pacifici}, Camilla and {Pascual}, Sergio and {Pascual-Granado}, J. and {Patil}, Rohit R. and {Perren}, Gabriel I. and {Pickering}, Timothy E. and {Rastogi}, Tanuj and {Roulston}, Benjamin R. and {Ryan}, Daniel F. and {Rykoff}, Eli S. and {Sabater}, Jose and {Sakurikar}, Parikshit and {Salgado}, Jes{'u}s and {Sanghi}, Aniket and {Saunders}, Nicholas and {Savchenko}, Volodymyr and {Schwardt}, Ludwig and {Seifert-Eckert}, Michael and {Shih}, Albert Y. and {Jain}, Anany Shrey and {Shukla}, Gyanendra and {Sick}, Jonathan and {Simpson}, Chris and {Singanamalla}, Sudheesh and {Singer}, Leo P. and {Singhal}, Jaladh and {Sinha}, Manodeep and {Sip{H{o}}cz}, Brigitta M. and {Spitler}, Lee R. and {Stansby}, David and {Streicher}, Ole and {{{S}}umak}, Jani and {Swinbank}, John D. and {Taranu}, Dan S. and {Tewary}, Nikita and {Tremblay}, Grant R. and {Val-Borro}, Miguel de and {Van Kooten}, Samuel J. and {Vasovi{'c}}, Zlatan and {Verma}, Shresth and {de Miranda Cardoso}, Jos{'e} Vin{'i}cius and {Williams}, Peter K.~G. and {Wilson}, Tom J. and {Winkel}, Benjamin and {Wood-Vasey}, W.~M. and {Xue}, Rui and {Yoachim}, Peter and {Zhang}, Chen and {Zonca}, Andrea and {Astropy Project Contributors}},
        title = "{The Astropy Project: Sustaining and Growing a Community-oriented Open-source Project and the Latest Major Release (v5.0) of the Core Package}",
      journal = {apj},
         year = 2022,
        month = aug,
       volume = {935},
       number = {2},
          eid = {167},
        pages = {167},
          doi = {10.3847/1538-4357/ac7c74},
archivePrefix = {arXiv},
       eprint = {2206.14220},
 primaryClass = {astro-ph.IM},
       adsurl = {https://ui.adsabs.harvard.edu/abs/2022ApJ...935..167A}
}

@misc{jax2018github,
  author = {Bradbury, James and Frostig, Roy and Hawkins, Peter and et al.},
  title = {{JAX}: composable transformations of Python+NumPy programs},
  year = {2018},
  howpublished = {\url{https://github.com/jax-ml/jax}}
}

@ARTICLE{emu,
       author = {{Norris}, Ray P. and {Hopkins}, A.~M. and {Afonso}, J. and {Brown}, S. and {Condon}, J.~J. and {Dunne}, L. and {Feain}, I. and {Hollow}, R. and {Jarvis}, M. and {Johnston-Hollitt}, M. and {Lenc}, E. and {Middelberg}, E. and {Padovani}, P. and {Prandoni}, I. and {Rudnick}, L. and {Seymour}, N. and {Umana}, G. and {Andernach}, H. and {Alexander}, D.~M. and {Appleton}, P.~N. and {Bacon}, D. and {Banfield}, J. and {Becker}, W. and {Brown}, M.~J.~I. and {Ciliegi}, P. and {Jackson}, C. and {Eales}, S. and {Edge}, A.~C. and {Gaensler}, B.~M. and {Giovannini}, G. and {Hales}, C.~A. and {Hancock}, P. and {Huynh}, M.~T. and {Ibar}, E. and {Ivison}, R.~J. and {Kennicutt}, R. and {Kimball}, Amy E. and {Koekemoer}, A.~M. and {Koribalski}, B.~S. and {L{\'o}pez-S{\'a}nchez}, {\'A}. R. and {Mao}, M.~Y. and {Murphy}, T. and {Messias}, H. and {Pimbblet}, K.~A. and {Raccanelli}, A. and {Randall}, K.~E. and {Reiprich}, T.~H. and {Roseboom}, I.~G. and {R{\"o}ttgering}, H. and {Saikia}, D.~J. and {Sharp}, R.~G. and {Slee}, O.~B. and {Smail}, Ian and {Thompson}, M.~A. and {Urquhart}, J.~S. and {Wall}, J.~V. and {Zhao}, G. -B.},
        title = "{EMU: Evolutionary Map of the Universe}",
      journal = {\pasa},
         year = 2011,
        month = aug,
       volume = {28},
       number = {3},
        pages = {215-248},
          doi = {10.1071/AS11021},
archivePrefix = {arXiv},
       eprint = {1106.3219},
 primaryClass = {astro-ph.CO},
       adsurl = {https://ui.adsabs.harvard.edu/abs/2011PASA...28..215N}
}

@ARTICLE{camera2015,
       author = {{Camera}, Stefano and {Santos}, M{\'a}rio G. and {Maartens}, Roy},
        title = "{Probing primordial non-Gaussianity with SKA galaxy redshift surveys: a fully relativistic analysis}",
      journal = {\mnras},
         year = 2015,
        month = apr,
       volume = {448},
       number = {2},
        pages = {1035-1043},
          doi = {10.1093/mnras/stv040},
archivePrefix = {arXiv},
       eprint = {1409.8286},
 primaryClass = {astro-ph.CO},
       adsurl = {https://ui.adsabs.harvard.edu/abs/2015MNRAS.448.1035C}
}

@article{Lacy:2015tha,
    author = "Lacy, Mark and Ridgway, Susan E. and Sajina, Anna and Petric, Andreea O. and Gates, Elinor L. and Urrutia, Tanya and Storrie-Lombardi, Lisa J.",
    title = "{The Spitzer mid-infrared AGN survey. II-the demographics and cosmic evolution of the AGN population}",
    eprint = "1501.04118",
    archivePrefix = "arXiv",
    primaryClass = "astro-ph.GA",
    doi = "10.1088/0004-637X/802/2/102",
    journal = "Astrophys. J.",
    volume = "802",
    number = "2",
    pages = "102",
    year = "2015"
}

@article{Bonaldi:2018xfm,
    author = "Bonaldi, Anna and Bonato, Matteo and Galluzzi, Vincenzo and Harrison, Ian and Massardi, Marcella and Kay, Scott and De Zotti, Gianfranco and Brown, Michael L.",
    title = "{The Tiered Radio Extragalactic Continuum Simulation (T-RECS)}",
    eprint = "1805.05222",
    archivePrefix = "arXiv",
    primaryClass = "astro-ph.GA",
    doi = "10.1093/mnras/sty2603",
    journal = "Mon. Not. Roy. Astron. Soc.",
    volume = "482",
    number = "1",
    pages = "2--19",
    year = "2019"
}

@ARTICLE{planck2020,
       author = {{Planck Collaboration} and {Aghanim}, N. and {Akrami}, Y. and {Arroja}, F. and {Ashdown}, M. and {Aumont}, J. and {Baccigalupi}, C. and {Ballardini}, M. and {Banday}, A.~J. and {Barreiro}, R.~B. and {Bartolo}, N. and {Basak}, S. and {Battye}, R. and {Benabed}, K. and {Bernard}, J. -P. and {Bersanelli}, M. and {Bielewicz}, P. and {Bock}, J.~J. and {Bond}, J.~R. and {Borrill}, J. and {Bouchet}, F.~R. and {Boulanger}, F. and {Bucher}, M. and {Burigana}, C. and {Butler}, R.~C. and {Calabrese}, E. and {Cardoso}, J. -F. and {Carron}, J. and {Casaponsa}, B. and {Challinor}, A. and {Chiang}, H.~C. and {Colombo}, L.~P.~L. and {Combet}, C. and {Contreras}, D. and {Crill}, B.~P. and {Cuttaia}, F. and {de Bernardis}, P. and {de Zotti}, G. and {Delabrouille}, J. and {Delouis}, J. -M. and {D{\'e}sert}, F. -X. and {Di Valentino}, E. and {Dickinson}, C. and {Diego}, J.~M. and {Donzelli}, S. and {Dor{\'e}}, O. and {Douspis}, M. and {Ducout}, A. and {Dupac}, X. and {Efstathiou}, G. and {Elsner}, F. and {En{\ss}lin}, T.~A. and {Eriksen}, H.~K. and {Falgarone}, E. and {Fantaye}, Y. and {Fergusson}, J. and {Fernandez-Cobos}, R. and {Finelli}, F. and {Forastieri}, F. and {Frailis}, M. and {Franceschi}, E. and {Frolov}, A. and {Galeotta}, S. and {Galli}, S. and {Ganga}, K. and {G{\'e}nova-Santos}, R.~T. and {Gerbino}, M. and {Ghosh}, T. and {Gonz{\'a}lez-Nuevo}, J. and {G{\'o}rski}, K.~M. and {Gratton}, S. and {Gruppuso}, A. and {Gudmundsson}, J.~E. and {Hamann}, J. and {Handley}, W. and {Hansen}, F.~K. and {Helou}, G. and {Herranz}, D. and {Hildebrandt}, S.~R. and {Hivon}, E. and {Huang}, Z. and {Jaffe}, A.~H. and {Jones}, W.~C. and {Karakci}, A. and {Keih{\"a}nen}, E. and {Keskitalo}, R. and {Kiiveri}, K. and {Kim}, J. and {Kisner}, T.~S. and {Knox}, L. and {Krachmalnicoff}, N. and {Kunz}, M. and {Kurki-Suonio}, H. and {Lagache}, G. and {Lamarre}, J. -M. and {Langer}, M. and {Lasenby}, A. and {Lattanzi}, M. and {Lawrence}, C.~R. and {Le Jeune}, M. and {Leahy}, J.~P. and {Lesgourgues}, J. and {Levrier}, F. and {Lewis}, A. and {Liguori}, M. and {Lilje}, P.~B. and {Lilley}, M. and {Lindholm}, V. and {L{\'o}pez-Caniego}, M. and {Lubin}, P.~M. and {Ma}, Y. -Z. and {Mac{\'\i}as-P{\'e}rez}, J.~F. and {Maggio}, G. and {Maino}, D. and {Mandolesi}, N. and {Mangilli}, A. and {Marcos-Caballero}, A. and {Maris}, M. and {Martin}, P.~G. and {Martinelli}, M. and {Mart{\'\i}nez-Gonz{\'a}lez}, E. and {Matarrese}, S. and {Mauri}, N. and {McEwen}, J.~D. and {Meerburg}, P.~D. and {Meinhold}, P.~R. and {Melchiorri}, A. and {Mennella}, A. and {Migliaccio}, M. and {Millea}, M. and {Mitra}, S. and {Miville-Desch{\^e}nes}, M. -A. and {Molinari}, D. and {Moneti}, A. and {Montier}, L. and {Morgante}, G. and {Moss}, A. and {Mottet}, S. and {M{\"u}nchmeyer}, M. and {Natoli}, P. and {N{\o}rgaard-Nielsen}, H.~U. and {Oxborrow}, C.~A. and {Pagano}, L. and {Paoletti}, D. and {Partridge}, B. and {Patanchon}, G. and {Pearson}, T.~J. and {Peel}, M. and {Peiris}, H.~V. and {Perrotta}, F. and {Pettorino}, V. and {Piacentini}, F. and {Polastri}, L. and {Polenta}, G. and {Puget}, J. -L. and {Rachen}, J.~P. and {Reinecke}, M. and {Remazeilles}, M. and {Renault}, C. and {Renzi}, A. and {Rocha}, G. and {Rosset}, C. and {Roudier}, G. and {Rubi{\~n}o-Mart{\'\i}n}, J.~A. and {Ruiz-Granados}, B. and {Salvati}, L. and {Sandri}, M. and {Savelainen}, M. and {Scott}, D. and {Shellard}, E.~P.~S. and {Shiraishi}, M. and {Sirignano}, C. and {Sirri}, G. and {Spencer}, L.~D. and {Sunyaev}, R. and {Suur-Uski}, A. -S. and {Tauber}, J.~A. and {Tavagnacco}, D. and {Tenti}, M. and {Terenzi}, L. and {Toffolatti}, L. and {Tomasi}, M. and {Trombetti}, T. and {Valiviita}, J. and {Van Tent}, B. and {Vibert}, L. and {Vielva}, P. and {Villa}, F. and {Vittorio}, N. and {Wandelt}, B.~D. and {Wehus}, I.~K. and {White}, M. and {White}, S.~D.~M. and {Zacchei}, A. and {Zonca}, A.},
        title = "{Planck 2018 results. I. Overview and the cosmological legacy of Planck}",
      journal = {\aap},
         year = 2020,
        month = sep,
       volume = {641},
          eid = {A1},
        pages = {A1},
          doi = {10.1051/0004-6361/201833880},
archivePrefix = {arXiv},
       eprint = {1807.06205},
 primaryClass = {astro-ph.CO},
       adsurl = {https://ui.adsabs.harvard.edu/abs/2020A&A...641A...1P}
}

@article{Bisigello_2021,
   title={Simulating the infrared sky with a SPRITZ},
   volume={651},
   ISSN={1432-0746},
   url={http://dx.doi.org/10.1051/0004-6361/202039909},
   DOI={10.1051/0004-6361/202039909},
   journal={Astronomy \&; Astrophysics},
   publisher={EDP Sciences},
   author={Bisigello, L. and Gruppioni, C. and Feltre, A. and Calura, F. and Pozzi, F. and Vignali, C. and Barchiesi, L. and Rodighiero, G. and Negrello, M.},
   year={2021},
   month=jul, pages={A52} }

@misc{emu-casda,
    author = {{Hopkins}, Andrew and {Norris}, Ray and {Vernstrom}, Tessa and {Kapinska}, Anna and {Marvil}, Josh},
    title = {ASKAP Data Products for Project AS201 (EMU): images and visibilities. v1.},
    year = {2022},
    publisher = {CSIRO},
    url = {ttp://hdl.handle.net/102.100.100/479788?index=1}
}

@article{Bonaldi:2023fnb,
    author = "Bonaldi, Anna and Hartley, Philippa and Ronconi, Tommaso and De Zotti, Gianfranco and Bonato, Matteo",
    title = "{The tiered radio extragalactic continuum (T-RECS) simulation II: H{\,}i emission and continuum-H{\,}i cross-correlation}",
    eprint = "2305.10175",
    archivePrefix = "arXiv",
    primaryClass = "astro-ph.GA",
    doi = "10.1093/mnras/stad1913",
    journal = "Mon. Not. Roy. Astron. Soc.",
    volume = "524",
    number = "1",
    pages = "993--1007",
    year = "2023"
}

@article{Euclid2025,
   title={Euclid preparation: LVII. Observational expectations for redshift z<7 active galactic nuclei in the Euclid Wide and Deep surveys},
   volume={693},
   ISSN={1432-0746},
   url={http://dx.doi.org/10.1051/0004-6361/202450894},
   DOI={10.1051/0004-6361/202450894},
   journal={Astronomy \&; Astrophysics},
   publisher={EDP Sciences},
   author={Selwood, M. and Fotopoulou, S. and Bremer, M. N. and Bisigello, L. and Landt, H. and Bañados, E. and Zamorani, G. and Shankar, F. and Stern, D. and Lusso, E. and Spinoglio, L. and Allevato, V. and Ricci, F. and Feltre, A. and Mannucci, F. and Salvato, M. and Bowler, R. A. A. and Mignoli, M. and Vergani, D. and La Franca, F. and Amara, A. and Andreon, S. and Auricchio, N. and Baldi, M. and Bardelli, S. and Bender, R. and Bodendorf, C. and Bonino, D. and Branchini, E. and Brescia, M. and Brinchmann, J. and Camera, S. and Capobianco, V. and Carbone, C. and Carretero, J. and Casas, S. and Castellano, M. and Cavuoti, S. and Cimatti, A. and Congedo, G. and Conselice, C. J. and Conversi, L. and Copin, Y. and Courbin, F. and Courtois, H. M. and Cropper, M. and Da Silva, A. and Degaudenzi, H. and Di Giorgio, A. M. and Dinis, J. and Dubath, F. and Dupac, X. and Dusini, S. and Farina, M. and Farrens, S. and Ferriol, S. and Frailis, M. and Franceschi, E. and Galeotta, S. and Gillis, B. and Giocoli, C. and Grazian, A. and Grupp, F. and Guzzo, L. and Haugan, S. V. H. and Hoekstra, H. and Holliman, M. S. and Holmes, W. and Hook, I. and Hormuth, F. and Hornstrup, A. and Hudelot, P. and Jahnke, K. and Keihänen, E. and Kermiche, S. and Kiessling, A. and Kubik, B. and Kümmel, M. and Kunz, M. and Kurki-Suonio, H. and Laureijs, R. and Ligori, S. and Lilje, P. B. and Lindholm, V. and Lloro, I. and Maino, D. and Maiorano, E. and Mansutti, O. and Marggraf, O. and Markovic, K. and Martinet, N. and Marulli, F. and Massey, R. and Medinaceli, E. and Mei, S. and Melchior, M. and Mellier, Y. and Meneghetti, M. and Merlin, E. and Meylan, G. and Moresco, M. and Moscardini, L. and Munari, E. and Niemi, S.-M. and Nightingale, J. W. and Padilla, C. and Paltani, S. and Pasian, F. and Pedersen, K. and Percival, W. J. and Pettorino, V. and Polenta, G. and Poncet, M. and Popa, L. A. and Pozzetti, L. and Raison, F. and Rebolo, R. and Renzi, A. and Rhodes, J. and Riccio, G. and Rix, H.-W. and Romelli, E. and Roncarelli, M. and Rossetti, E. and Saglia, R. and Sapone, D. and Sartoris, B. and Scaramella, R. and Schirmer, M. and Schneider, P. and Schrabback, T. and Scialpi, M. and Secroun, A. and Seidel, G. and Serrano, S. and Sirignano, C. and Sirri, G. and Stanco, L. and Surace, C. and Tallada-Crespí, P. and Tavagnacco, D. and Taylor, A. N. and Teplitz, H. I. and Tereno, I. and Toledo-Moreo, R. and Torradeflot, F. and Tutusaus, I. and Valenziano, L. and Vassallo, T. and Veropalumbo, A. and Wang, Y. and Weller, J. and Zucca, E. and Biviano, A. and Bolzonella, M. and Bozzo, E. and Burigana, C. and Colodro-Conde, C. and De Lucia, G. and Di Ferdinando, D. and Escartin Vigo, J. A. and Farinelli, R. and George, K. and Gracia-Carpio, J. and Martinelli, M. and Mauri, N. and Neissner, C. and Sakr, Z. and Scottez, V. and Tenti, M. and Viel, M. and Wiesmann, M. and Akrami, Y. and Anselmi, S. and Baccigalupi, C. and Ballardini, M. and Bethermin, M. and Blanchard, A. and Blot, L. and Borgani, S. and Bruton, S. and Cabanac, R. and Calabro, A. and Canas-Herrera, G. and Cappi, A. and Carvalho, C. S. and Castignani, G. and Castro, T. and Chambers, K. C. and Contarini, S. and Contini, T. and Cooray, A. R. and Cucciati, O. and Davini, S. and De Caro, B. and Desprez, G. and Díaz-Sánchez, A. and Di Domizio, S. and Dole, H. and Escoffier, S. and Ferrari, A. G. and Ferrero, I. and Finelli, F. and Fontana, A. and Fornari, F. and Gabarra, L. and Ganga, K. and García-Bellido, J. and Gautard, V. and Gaztanaga, E. and Giacomini, F. and Gozaliasl, G. and Hall, A. and Hildebrandt, H. and Hjorth, J. and Kajava, J. J. E. and Kansal, V. and Karagiannis, D. and Kirkpatrick, C. C. and Legrand, L. and Libet, G. and Loureiro, A. and Macias-Perez, J. and Maggio, G. and Magliocchetti, M. and Maoli, R. and Martins, C. J. A. P. and Matthew, S. and Maurin, L. and Metcalf, R. B. and Monaco, P. and Moretti, C. and Morgante, G. and Nadathur, S. and Nicastro, L. and Walton, N. A. and Patrizii, L. and Pezzotta, A. and Pöntinen, M. and Popa, V. and Porciani, C. and Potter, D. and Risso, I. and Rocci, P.-F. and Sahlén, M. and Sánchez, A. G. and Schneider, A. and Sefusatti, E. and Sereno, M. and Simon, P. and Spurio Mancini, A. and Steinwagner, J. and Testera, G. and Teyssier, R. and Toft, S. and Tosi, S. and Troja, A. and Tucci, M. and Valieri, C. and Valiviita, J. and Verza, G. and Weaver, J. R. and Zinchenko, I. A.},
   year={2025},
   month=jan, pages={A250} }

@book{milne1935,
  title     = {Relativity, Gravitation and World-structure},
  publisher = {Oxford University Press},
  year      = {1935},
  author    = {Milne, Edward A.},
}

@book{Weinberg:2008zzc,
    author = "Weinberg, Steven",
    title = "{Cosmology}",
    year = "2008",
    publisher = "Oxford University Press"
}

@ARTICLE{blake2002a,
       author = {{Blake}, Chris and {Wall}, Jasper},
        title = "{Measurement of the angular correlation function of radio galaxies from the NRAO VLA Sky Survey}",
      journal = {\mnras},
         year = 2002,
        month = jan,
       volume = {329},
       number = {2},
        pages = {L37-L41},
          doi = {10.1046/j.1365-8711.2002.05163.x},
archivePrefix = {arXiv},
       eprint = {astro-ph/0111328},
 primaryClass = {astro-ph},
       adsurl = {https://ui.adsabs.harvard.edu/abs/2002MNRAS.329L..37B}
}

@article{Martin:2025ywz,
    author = "Mart{\'\i}n, Alicia and Skordis, Constantinos and Bartlett, Deaglan J. and Desmond, Harry and Ferreira, Pedro G. and Yasin, Tariq",
    title = "{The Cosmological Dipole in Tilted Anisotropic Universes}",
    eprint = "2512.03867",
    archivePrefix = "arXiv",
    primaryClass = "astro-ph.CO",
    month = "12",
    year = "2025"
}

@article{Blumke:2025nrq,
    author = {Bl{\"u}mke, Maximilian and Schmitz, Kai and Schr{\"o}der, Tobias and Agarwal, Deepali and Romano, Joseph D.},
    title = "{Kinematic Anisotropies in PTA Observations: Analytical Toolkit}",
    eprint = "2512.24055",
    archivePrefix = "arXiv",
    primaryClass = "gr-qc",
    reportNumber = "MS-TP-25-39",
    doi = "10.3390/sym18020355",
    journal = "Symmetry",
    volume = "18",
    pages = "355",
    year = "2026"
}

@article{Yasin:2026nte,
    author = "Yasin, Tariq and Stiskalek, Richard and Desmond, Harry and von Hausegger, Sebastian and Ferreira, Pedro G.",
    title = "{Testing cosmic anisotropy with cluster scaling relations}",
    eprint = "2602.06007",
    archivePrefix = "arXiv",
    primaryClass = "astro-ph.CO",
    month = "2",
    year = "2026"
}

@article{Millon:2026nwo,
    author = "Millon, Martin and Dalang, Charles and Collett, Thomas and Bonvin, Camille",
    title = "{Kinematic cosmic dipole from a large sample of strong lenses}",
    eprint = "2603.11152",
    archivePrefix = "arXiv",
    primaryClass = "astro-ph.CO",
    month = "3",
    year = "2026"
}

@article{Peebles_2022,
	doi = {10.1016/j.aop.2022.169159},
	url = {https://doi.org/10.1016%2Fj.aop.2022.169159},
	year = 2022,
	month = dec,
	publisher = {Elsevier {BV}},
	volume = {447},
	pages = {169159},
	author = {P.J.E. Peebles},
	title = {Anomalies in physical cosmology},
	journal = {Annals of Physics}
}

@article{dalang2022,
    author = {Dalang, Charles and Bonvin, Camille},
    title = "{On the kinematic cosmic dipole tension}",
    journal = {Monthly Notices of the Royal Astronomical Society},
    volume = {512},
    number = {3},
    pages = {3895-3905},
    year = {2022},
    month = {03},
    issn = {0035-8711},
    doi = {10.1093/mnras/stac726},
    url = {https://doi.org/10.1093/mnras/stac726},
    eprint = {https://academic.oup.com/mnras/article-pdf/512/3/3895/43290836/stac726.pdf},
}

@ARTICLE{aluri2023,
       author = {{Kumar Aluri}, Pavan and {Cea}, Paolo and {Chingangbam}, Pravabati and {Chu}, Ming-Chung and {Clowes}, Roger G. and {Hutsem{\'e}kers}, Damien and {Kochappan}, Joby P. and {Lopez}, Alexia M. and {Liu}, Lang and {Martens}, Niels C.~M. and {Martins}, C.~J.~A.~P. and {Migkas}, Konstantinos and {{\'O} Colg{\'a}in}, Eoin and {Pranav}, Pratyush and {Shamir}, Lior and {Singal}, Ashok K. and {Sheikh-Jabbari}, M.~M. and {Wagner}, Jenny and {Wang}, Shao-Jiang and {Wiltshire}, David L. and {Yeung}, Shek and {Yin}, Lu and {Zhao}, Wen},
        title = "{Is the observable Universe consistent with the cosmological principle?}",
      journal = {Classical and Quantum Gravity},
         year = 2023,
        month = may,
       volume = {40},
       number = {9},
          eid = {094001},
        pages = {094001},
          doi = {10.1088/1361-6382/acbefc},
archivePrefix = {arXiv},
       eprint = {2207.05765},
 primaryClass = {astro-ph.CO},
       adsurl = {https://ui.adsabs.harvard.edu/abs/2023CQGra..40i4001K}
}

@article{Secrest2025,
	author = {Secrest, Nathan J. and von Hausegger, Sebastian and Rameez, Mohamed and Mohayaee, Roya and Sarkar, Subir},
        month = jan,
	doi = {10.1038/s42254-024-00803-3},
	journal = {Nature Reviews Physics},
	title = {Forty years of the Ellis--Baldwin test},
	url = {https://doi.org/10.1038/s42254-024-00803-3},
	year = {2025}}

@article{Secrest:2025wyu,
    author = "Secrest, Nathan and von Hausegger, Sebastian and Rameez, Mohamed and Mohayaee, Roya and Sarkar, Subir",
    title = "{Colloquium: The Cosmic Dipole Anomaly}",
    eprint = "2505.23526",
    archivePrefix = "arXiv",
    primaryClass = "astro-ph.CO",
    month = "5",
    year = "2025"
}

@ARTICLE{ellis1984,
       author = {{Ellis}, G.~F.~R. and {Baldwin}, J.~E.},
        title = "{On the expected anisotropy of radio source counts}",
      journal = {\mnras},
         year = 1984,
        month = jan,
       volume = {206},
        pages = {377-381},
          doi = {10.1093/mnras/206.2.377},
       adsurl = {https://ui.adsabs.harvard.edu/abs/1984MNRAS.206..377E}
}

@ARTICLE{2002Natur.416..150B,
       author = {{Blake}, Chris and {Wall}, Jasper},
        title = "{A velocity dipole in the distribution of radio galaxies}",
      journal = {\nat},
         year = 2002,
        month = mar,
       volume = {416},
       number = {6877},
        pages = {150-152},
          doi = {10.1038/416150a},
archivePrefix = {arXiv},
       eprint = {astro-ph/0203385},
 primaryClass = {astro-ph},
       adsurl = {https://ui.adsabs.harvard.edu/abs/2002Natur.416..150B}
}

@article{Crawford_2009,
   title={DETECTING THE COSMIC DIPOLE ANISOTROPY IN LARGE-SCALE RADIO SURVEYS},
   volume={692},
   ISSN={1538-4357},
   url={http://dx.doi.org/10.1088/0004-637X/692/1/887},
   DOI={10.1088/0004-637x/692/1/887},
   number={1},
   journal={The Astrophysical Journal},
   publisher={American Astronomical Society},
   author={Crawford, Fronefield},
   year={2009},
   month=feb, pages={887–893}
}

@article{singal2011,
    doi = {10.1088/2041-8205/742/2/L23},
    url = {https://dx.doi.org/10.1088/2041-8205/742/2/L23},
    year = {2011},
    month = nov,
    publisher = {The American Astronomical Society},
    volume = {742},
    number = {2},
    pages = {L23},
    author = {Ashok K. Singal},
    title = {LARGE PECULIAR MOTION OF THE SOLAR SYSTEM FROM THE DIPOLE ANISOTROPY IN SKY BRIGHTNESS DUE TO DISTANT RADIO SOURCES},
    journal = {The Astrophysical Journal Letters}
}

@article{rubart2013,
	doi = {10.1051/0004-6361/201321215},
	url = {https://doi.org/10.1051%2F0004-6361%2F201321215},
	year = 2013,
	month = jul,
	publisher = {{EDP} Sciences},
	volume = {555},
	pages = {A117},
	author = {M. {Rubart} and D. J. {Schwarz}},
	title = {Cosmic radio dipole from {NVSS} and {WENSS}},
	journal = {\aap}
}

@article{tiwari2016,
    doi = {10.1088/1475-7516/2016/03/062},
    url = {https://dx.doi.org/10.1088/1475-7516/2016/03/062},
    year = {2016},
    month = {mar},
    publisher = {},
    volume = {2016},
    number = {03},
    pages = {062},
    author = {Prabhakar Tiwari and Adi Nusser},
    title = {Revisiting the NVSS number count dipole},
    journal = {Journal of Cosmology and Astroparticle Physics}
}

@article{colin2017,
  author        = {{Colin}, Jacques and {Mohayaee}, Roya and {Rameez}, Mohamed and {Sarkar}, Subir},
  title         = {{High-redshift radio galaxies and divergence from the CMB dipole}},
  journal       = {\mnras},
  year          = 2017,
  month         = oct,
  volume        = {471},
  number        = {1},
  pages         = {1045-1055},
  doi           = {10.1093/mnras/stx1631},
  archiveprefix = {arXiv},
  eprint        = {1703.09376},
  primaryclass  = {astro-ph.CO},
  adsurl        = {https://ui.adsabs.harvard.edu/abs/2017MNRAS.471.1045C}
}

@article{bengaly2018,
  doi       = {10.1088/1475-7516/2018/04/031},
  url       = {https://doi.org/10.1088/1475-7516/2018/04/031},
  year      = 2018,
  month     = apr,
  publisher = {{IOP} Publishing},
  volume    = {2018},
  number    = {04},
  pages     = {031--031},
  author    = {Carlos A.P. Bengaly and Roy Maartens and Mario G. Santos},
  title     = {Probing the Cosmological Principle in the counts of radio galaxies at different frequencies},
  journal   = {\jcap}
}

@ARTICLE{secrest2021,
       author = {{Secrest}, Nathan J. and {von Hausegger}, Sebastian and {Rameez}, Mohamed and {Mohayaee}, Roya and {Sarkar}, Subir and {Colin}, Jacques},
        title = "{A Test of the Cosmological Principle with Quasars}",
      journal = {\apjl},
         year = 2021,
        month = feb,
       volume = {908},
       number = {2},
          eid = {L51},
        pages = {L51},
          doi = {10.3847/2041-8213/abdd40},
archivePrefix = {arXiv},
       eprint = {2009.14826},
 primaryClass = {astro-ph.CO},
       adsurl = {https://ui.adsabs.harvard.edu/abs/2021ApJ...908L..51S}
}

@article{siewert2021,
  author  = {{Siewert}, Thilo M. and {Schmidt-Rubart}, Matthias and {Schwarz}, Dominik J.},
  title   = {Cosmic radio dipole: Estimators and frequency dependence},
  doi     = {10.1051/0004-6361/202039840},
  url     = {https://doi.org/10.1051/0004-6361/202039840},
  journal = {\aap},
  year    = 2021,
  volume  = 653,
  pages   = {A9}
}

@ARTICLE{darling2022,
       author = {{Darling}, Jeremy},
        title = "{The Universe is Brighter in the Direction of Our Motion: Galaxy Counts and Fluxes are Consistent with the CMB Dipole}",
      journal = {\apjl},
         year = 2022,
        month = jun,
       volume = {931},
       number = {2},
          eid = {L14},
        pages = {L14},
          doi = {10.3847/2041-8213/ac6f08},
archivePrefix = {arXiv},
       eprint = {2205.06880},
 primaryClass = {astro-ph.CO},
       adsurl = {https://ui.adsabs.harvard.edu/abs/2022ApJ...931L..14D}
}

@ARTICLE{2022ApJ...937L..31S,
       author = {{Secrest}, Nathan J. and {von Hausegger}, Sebastian and {Rameez}, Mohamed and {Mohayaee}, Roya and {Sarkar}, Subir},
        title = "{A Challenge to the Standard Cosmological Model}",
      journal = {\apjl},
         year = 2022,
        month = oct,
       volume = {937},
       number = {2},
          eid = {L31},
        pages = {L31},
          doi = {10.3847/2041-8213/ac88c0},
archivePrefix = {arXiv},
       eprint = {2206.05624},
 primaryClass = {astro-ph.CO},
       adsurl = {https://ui.adsabs.harvard.edu/abs/2022ApJ...937L..31S}
}

@article{wagenveld2023,
	author = {{Wagenveld}, J. D. and {Kl\"ockner}, H.-R. and {Schwarz}, D. J.},
	title = {The cosmic radio dipole: Bayesian estimators on new and old radio surveys},
	DOI= "10.1051/0004-6361/202346210",
	url= "https://doi.org/10.1051/0004-6361/202346210",
	journal = {A\&A},
	year = 2023,
	volume = 675,
	pages = "A72",
}

@article{vonHausegger:2024jan,
    author = "von Hausegger, Sebastian",
    title = "{The expected kinematic matter dipole is robust against source evolution}",
    eprint = "2404.07929",
    archivePrefix = "arXiv",
    primaryClass = "astro-ph.CO",
    doi = "10.1093/mnrasl/slae092",
    journal = "Mon. Not. Roy. Astron. Soc.",
    volume = "535",
    number = "1",
    pages = "L49--L53",
    year = "2024"
}

@article{Bohme:2025nvu,
    author = {B{\"o}hme, Lukas and Schwarz, Dominik J. and Tiwari, Prabhakar and Pashapour-Ahmadabadi, Morteza and Bahr-Kalus, Benedict and Bilicki, Maciej and Hale, Catherine L. and Heneka, Caroline S. and Siewert, Thilo M.},
    title = "{Overdispersed Radio Source Counts and Excess Radio Dipole Detection}",
    eprint = "2509.16732",
    archivePrefix = "arXiv",
    primaryClass = "astro-ph.CO",
    doi = "10.1103/6z32-3zf4",
    journal = "Phys. Rev. Lett.",
    volume = "135",
    number = "20",
    pages = "201001",
    year = "2025"
}

@article{vonHausegger:2025iuo,
    author = "von Hausegger, Sebastian and Secrest, Nathan and Desmond, Harry and Rameez, Mohamed and Mohayaee, Roya and Sarkar, Subir",
    title = "{Clustering properties of the CatWISE2020 quasar catalogue and their impact on the cosmic dipole anomaly}",
    eprint = "2510.23769",
    archivePrefix = "arXiv",
    primaryClass = "astro-ph.CO",
    doi = "10.1093/mnras/stag201",
    journal = "Mon. Not. Roy. Astron. Soc.",
    volume = "546",
    number = "4",
    pages = "stag201",
    year = "2026"
}

@article{Takeuchi:2026can,
    author = "Takeuchi, Tsutomu T.",
    title = "{A General Formulation of the Kinematic Dipole as a Functional of Selection and Source Properties: Beyond the Ellis--Baldwin Approximation}",
    eprint = "2602.07389",
    archivePrefix = "arXiv",
    primaryClass = "astro-ph.CO",
    month = "2",
    year = "2026"
}

@article{Bonnefous:2026hpe,
    author = "Bonnefous, Albert",
    title = "{The kinematic cosmic dipole beyond Ellis and Baldwin}",
    eprint = "2602.05700",
    archivePrefix = "arXiv",
    primaryClass = "astro-ph.CO",
    month = "2",
    year = "2026"
}

@ARTICLE{dam2023,
       author = {{Dam}, Lawrence and {Lewis}, Geraint F. and {Brewer}, Brendon J.},
        title = "{Testing the cosmological principle with CatWISE quasars: a bayesian analysis of the number-count dipole}",
      journal = {\mnras},
         year = 2023,
        month = oct,
       volume = {525},
       number = {1},
        pages = {231-245},
          doi = {10.1093/mnras/stad2322},
archivePrefix = {arXiv},
       eprint = {2212.07733},
 primaryClass = {astro-ph.CO},
       adsurl = {https://ui.adsabs.harvard.edu/abs/2023MNRAS.525..231D}
}

@ARTICLE{mittal2024,
       author = {{Mittal}, Vasudev and {Oayda}, Oliver T. and {Lewis}, Geraint F.},
        title = "{The cosmic dipole in the Quaia sample of quasars: a Bayesian analysis}",
      journal = {\mnras},
         year = 2024,
        month = jan,
       volume = {527},
       number = {3},
        pages = {8497-8510},
          doi = {10.1093/mnras/stad3706},
archivePrefix = {arXiv},
       eprint = {2311.14938},
 primaryClass = {astro-ph.CO},
       adsurl = {https://ui.adsabs.harvard.edu/abs/2024MNRAS.527.8497M}
}

@article{Oayda:2024hnu,
    author = "Oayda, Oliver T. and Mittal, Vasudev and Lewis, Geraint F. and Murphy, Tara",
    title = "{A Bayesian approach to the cosmic dipole in radio galaxy surveys: joint analysis of NVSS~\& RACS}",
    eprint = "2406.01871",
    archivePrefix = "arXiv",
    primaryClass = "astro-ph.CO",
    doi = "10.1093/mnras/stae1399",
    journal = "Mon. Not. Roy. Astron. Soc.",
    volume = "531",
    number = "4",
    pages = "4545--4559",
    year = "2024"
}

@article{Oayda:2024voo,
    author = "Oayda, Oliver T. and Mittal, Vasudev and Lewis, Geraint F.",
    title = "{Cosmic multipoles in galaxy surveys \textendash{} I. How inferences depend on source counts and masks}",
    eprint = "2412.12600",
    archivePrefix = "arXiv",
    primaryClass = "astro-ph.CO",
    doi = "10.1093/mnras/stae2776",
    journal = "Mon. Not. Roy. Astron. Soc.",
    volume = "537",
    number = "1",
    pages = "1--20",
    year = "2025"
}

@article{Land-Strykowski:2025gkz,
    author = "Land-Strykowski, Mali and Lewis, Geraint F. and Murphy, Tara",
    title = "{Cosmic dipole tensions: confronting the cosmic microwave background with infrared and radio populations of cosmological sources}",
    eprint = "2509.18689",
    archivePrefix = "arXiv",
    primaryClass = "astro-ph.CO",
    doi = "10.1093/mnras/staf1621",
    journal = "Mon. Not. Roy. Astron. Soc.",
    volume = "543",
    number = "4",
    pages = "3229--3241",
    year = "2025"
}

@article{Oayda:2026afy,
    author = "Oayda, Oliver T. and Lewis, Geraint F.",
    title = "{Wising up to CatWISE: using simulation-based inference to interpret the ecliptic bias and confirm the cosmic dipole excess}",
    eprint = "2602.05070",
    archivePrefix = "arXiv",
    primaryClass = "astro-ph.CO",
    doi = "10.1093/mnras/stag248",
    journal = "Mon. Not. Roy. Astron. Soc.",
    volume = "546",
    number = "4",
    pages = "stag248",
    year = "2026"
}

\newpage
\appendix
\section{Posterior distributions for the joint analyses}
\begin{figure*}[!h]
\centering
\includegraphics[width=0.88\linewidth]{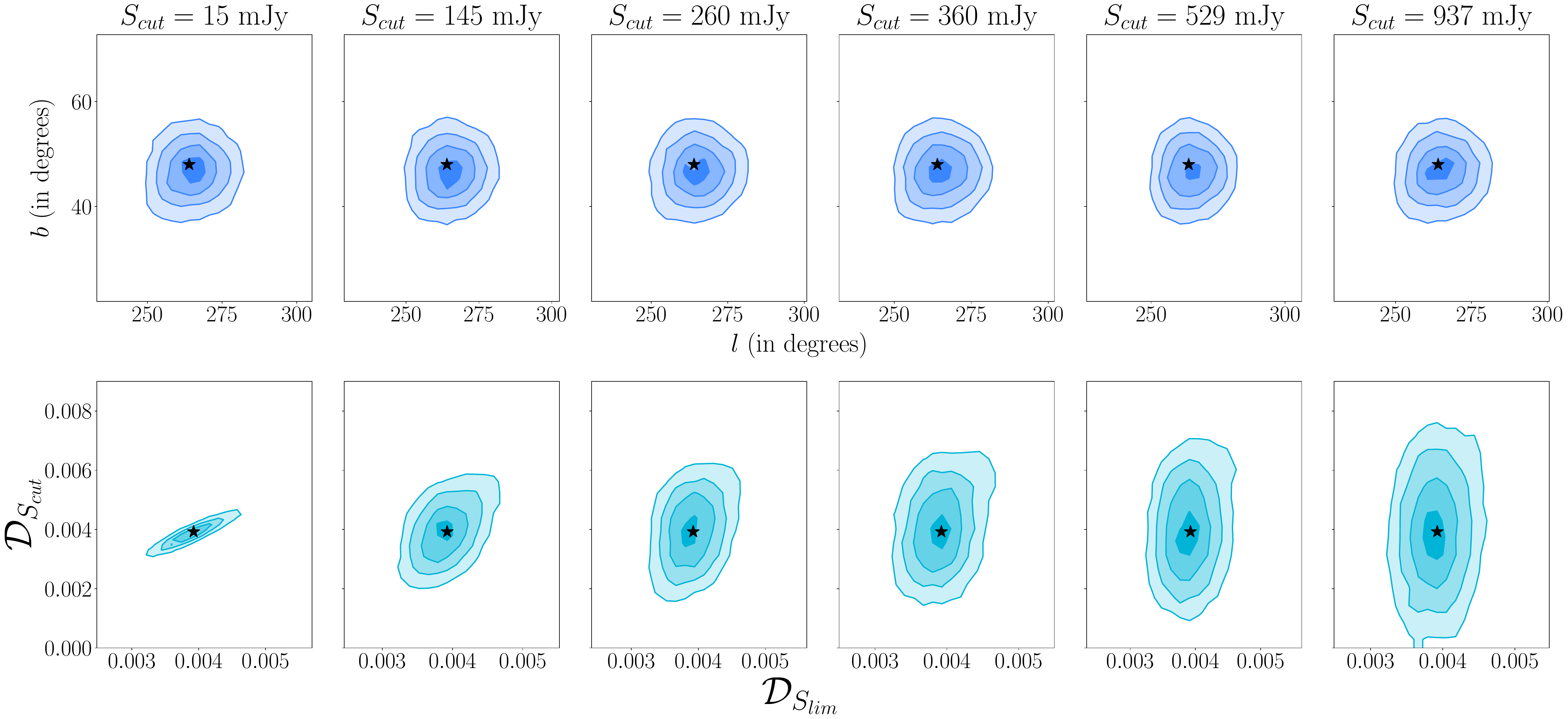}
\caption{
        Evolution of the consolidated posterior distributions of the dipole parameters for two-bin joint analysis of the mock catalogues constructed using the power-law luminosity function with $x=0.7$ and $\alpha=0.7$.
        In each column, the top histogram gives the distribution of the inferred dipole directions, while the bottom histogram gives the distribution for the inferred dipole amplitudes at both the lower flux limit $S_{\texttt{lim}}$ and the flux cut $S_{\texttt{cut}}$.
        The title of each column indicates the magnitude of the flux cut.
        The true parameter values are marked in black.
        The contours enclose $12\%$, $39\%$, $68\%$ and $86\%$ of the distribution in the 2D histograms.
        }
\label{fig: A1}
\end{figure*}
\vspace{-5mm}
\begin{figure*}[!h]
\centering
\includegraphics[width=0.88\linewidth]{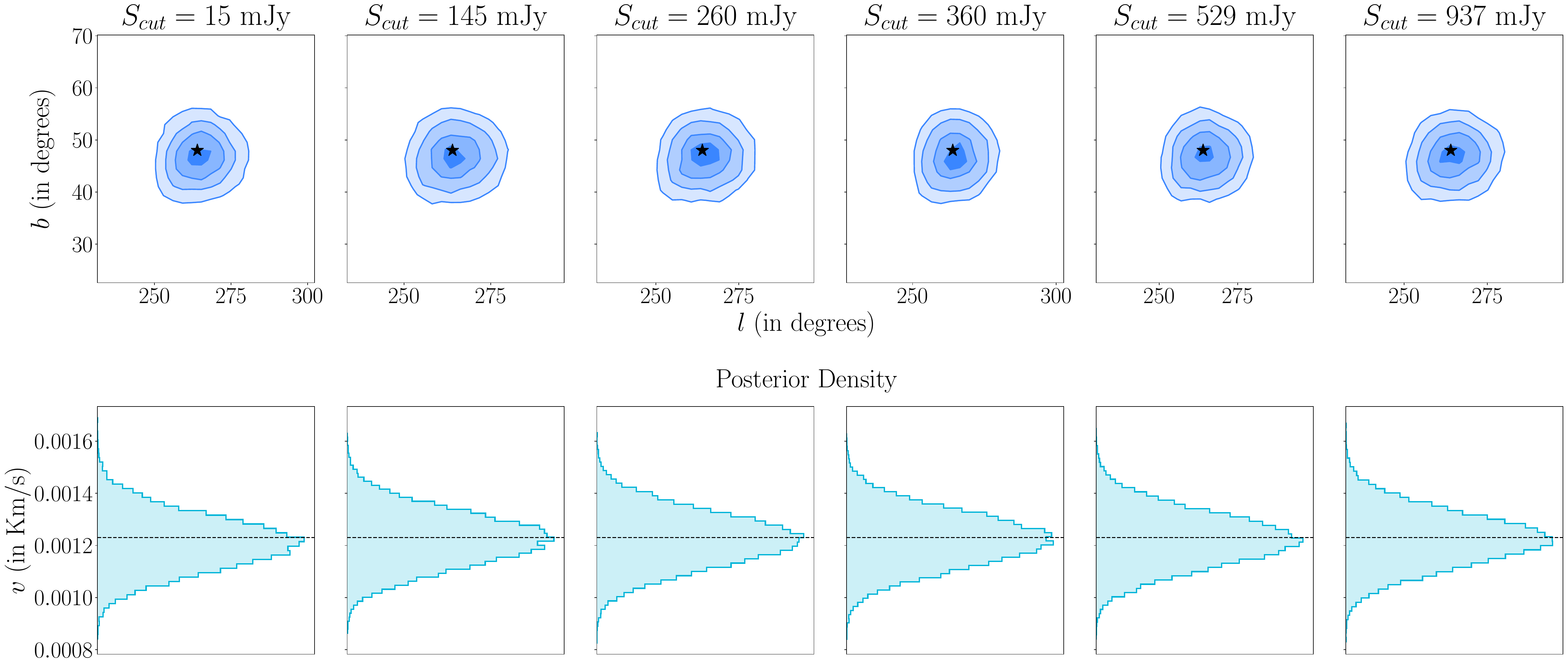}
\caption{
        Evolution of the consolidated posterior distributions of the dipole parameters for two-bin joint analysis of the mock catalogues constructed using the power-law luminosity function with $x=0.7$ and $\alpha=0.7$.
        In each column, the 2D histogram gives the distribution of the inferred dipole directions, while the 1D histogram gives the distribution for fitting the dipole velocity at both the lower flux limit $S_{\texttt{lim}}$ and the flux cut $S_{\texttt{cut}}$.
        The title of each column indicates the magnitude of the flux cut.
        The true parameter values are marked in black.
        The contours enclose $12\%$, $39\%$, $68\%$ and $86\%$ of the distribution in the 2D histograms.
        }
\label{fig: A2}
\end{figure*}
\vspace{-5mm}
\begin{figure*}[!h]
\centering
\includegraphics[width=0.88\linewidth]{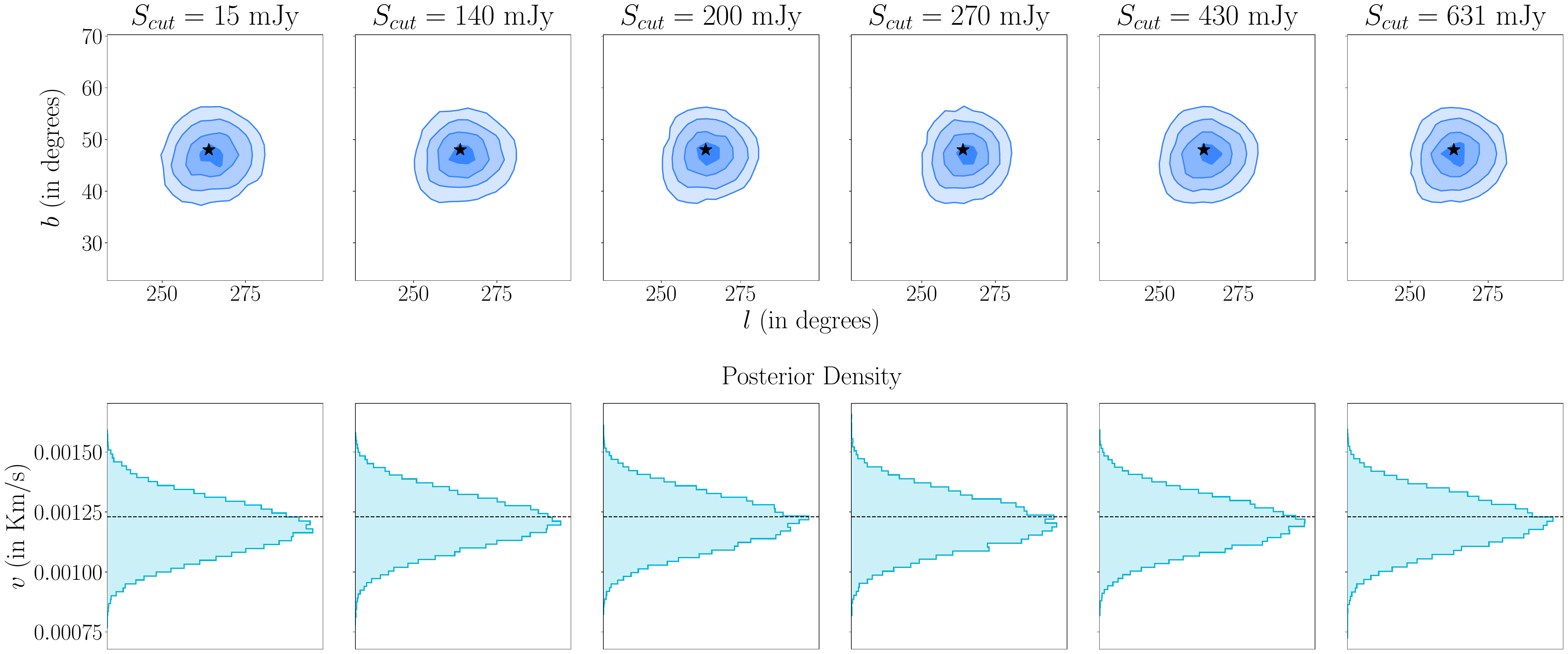}
\caption{As in Figure \ref{fig: A2}, but for the catalogues constructed using double power-law LF with $x_2 = 1.4$ and $\alpha=0.9$.}
\label{fig: A3}
\end{figure*}
\vspace{-5mm}
\begin{figure*}[!h]
\centering
\includegraphics[width=0.88\linewidth]{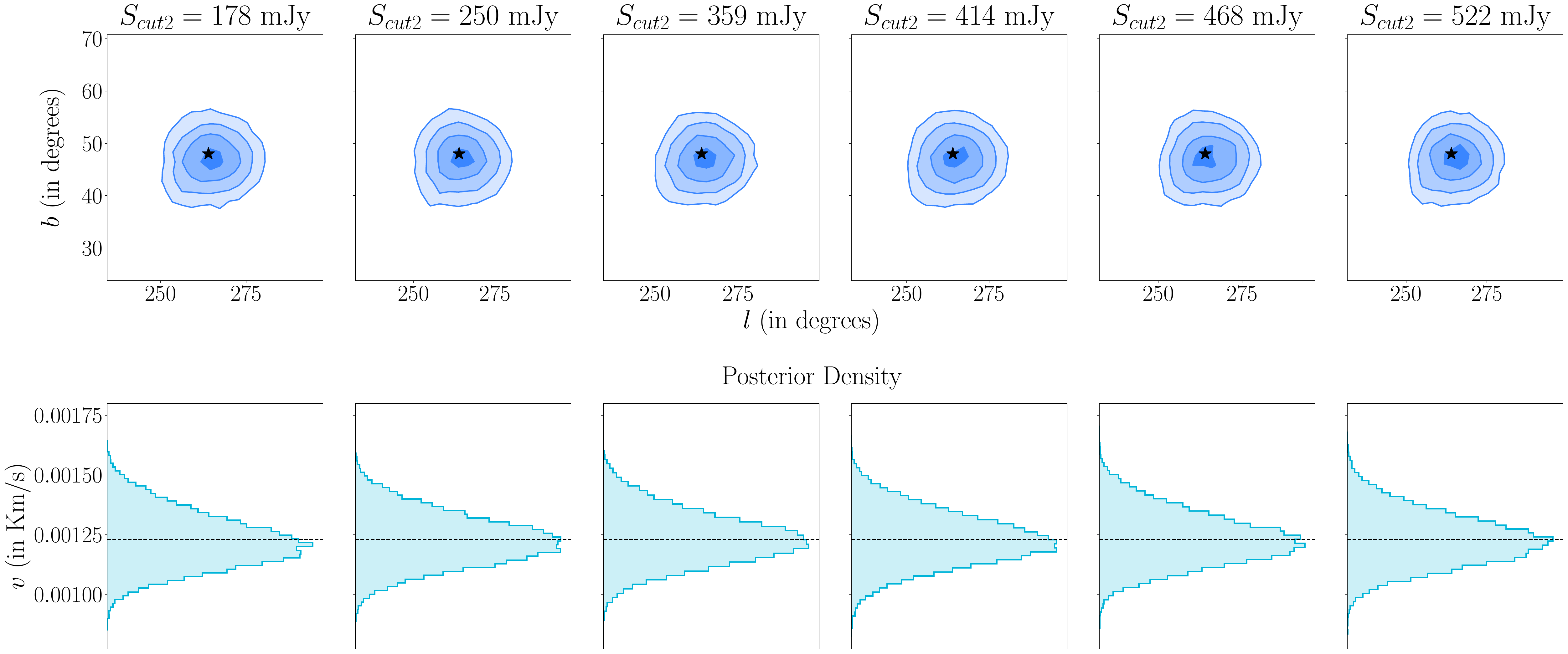}
\caption{As in Figure \ref{fig: A2}, but for the three bin joint analysis of catalogues constructed using double power-law LF with $x_2 = 1.7$ and $\alpha=0.7$.
        The first flux cut is fixed at $141\ mJy$, while the title of each column indicates the magnitude of the second flux cut.
        }
\label{fig: A4}
\end{figure*}
\vspace{-5mm}
\begin{figure*}[!h]
\centering
\includegraphics[width=0.88\linewidth]{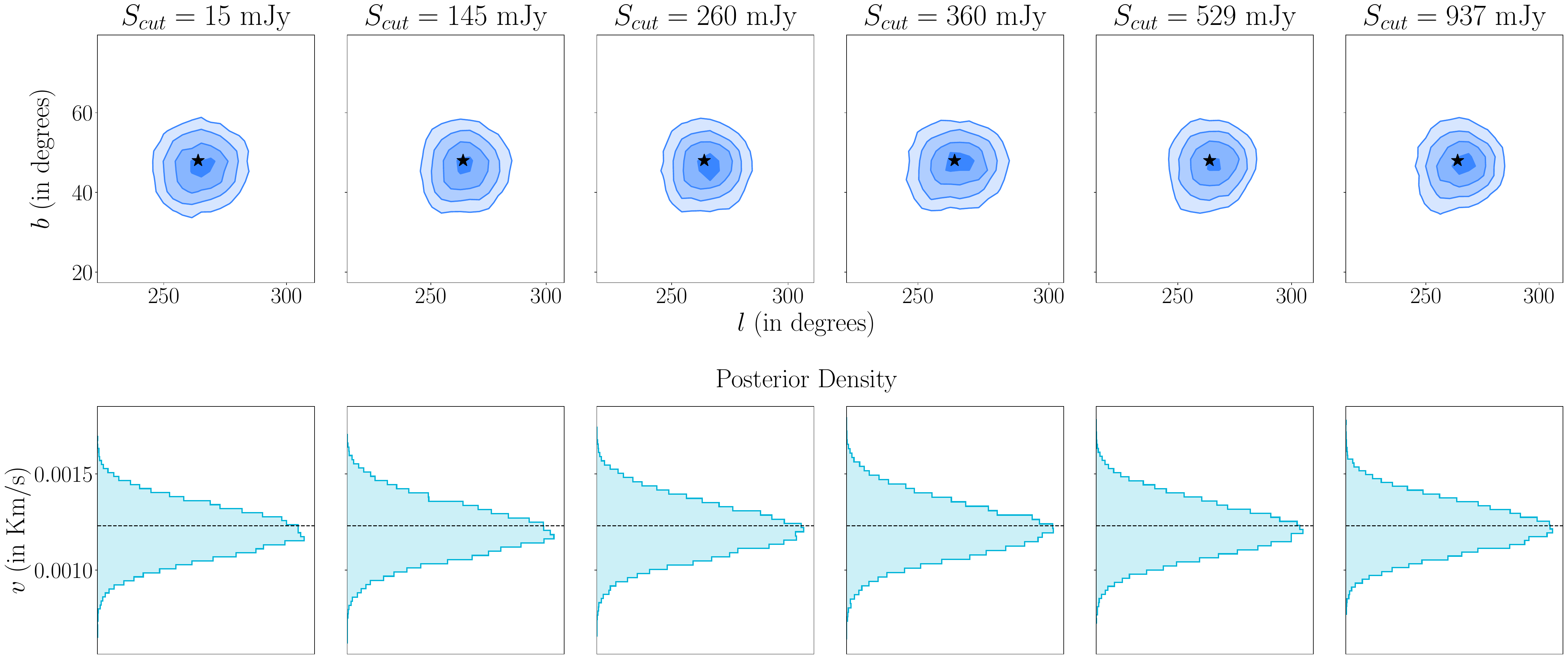}
\caption{As in Figure \ref{fig: A2}, but for the catalogues constructed using Schechter LF with $\upsilon=0.9$ and $\alpha=0.7$.}
\label{fig: A5}
\end{figure*}
\vspace{-5mm}
\begin{figure*}[!h]
\centering
\includegraphics[width=0.88\linewidth]{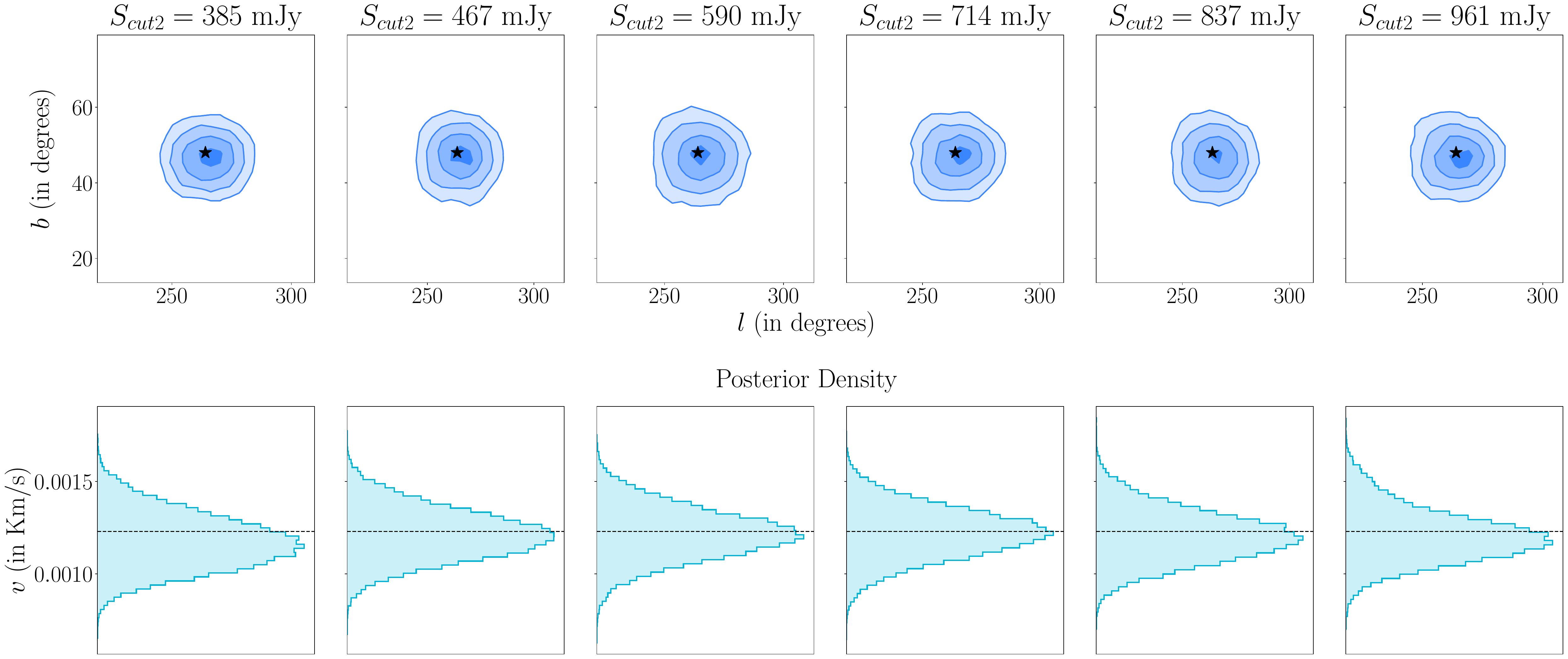}
\caption{As in Figure \ref{fig: A4}, but for the catalogues constructed using Schechter LF with $\upsilon=0.7$ and $\alpha=0.7$.
        The first flux cut is fixed at $344\ mJy$, while the title of each column indicates the magnitude of the second flux cut.}
\label{fig: A6}
\end{figure*}

\end{document}